\documentclass[12pt,preprint]{emulateapj}
\usepackage{graphicx}

\newcommand{\be}{\begin{equation}}
  \newcommand{\bel}[1]{\be\label{eq:#1}}
  \newcommand{\ee}{\end{equation}} \newcommand{\ba}{\begin{eqnarray}}
  \newcommand{\ea}{\end{eqnarray}} \newcommand\bp{\begin{figure}}
  \newcommand\ep{\end{figure}} \newcommand\bpm{\begin{figure*}}
  \newcommand\epm{\end{figure*}} \newcommand{\btab}{\begin{tabular}}
  \newcommand{\etab}{\end{tabular}} \newcommand{\bt}{\begin{table}}
  \newcommand{\et}{\end{table}} \newcommand{\ben}{\begin{enumerate}}
  \newcommand{\een}{\end{enumerate}} \newcommand\reffig[1]{Figure
  \ref{fig:#1}} \newcommand\refeq[1]{Equation \ref{eq:#1}}
\newcommand\refsec[1]{\S \ref{sec:#1}} \newcommand\reftbl[1]{Table
  \ref{tbl:#1}}

\newcommand{\bcn}{\begin{center}}
\newcommand{\ecn}{\end{center}}

\newcommand\Res{{\rm\bf r}}
\newcommand{\rmbf}[1]{{\rm\bf #1}}
\newcommand{\hal}{H$\alpha$}

\begin{document}

\title{Extended Anomalous Foreground Emission in the WMAP 3-Year Data}

\author{Gregory Dobler\altaffilmark{1,2} \& Douglas P. 
Finkbeiner\altaffilmark{1}}

\altaffiltext{1}{
Institute for Theory and Computation, 
Harvard-Smithsonian Center for Astrophysics, 60 Garden Street, MS-51,
Cambridge, MA 02138 USA
}
\altaffiltext{2}{gdobler@cfa.harvard.edu}

\begin{abstract}
  We study the spectral and morphological characteristics of the
  diffuse Galactic emission in the WMAP temperature data using a
  template-based multi-linear regression, and obtain the following
  results.  1. We confirm previous observations of a bump in the
  dust-correlated spectrum, consistent with the Draine \& Lazarian
  spinning dust model.  2. We also confirm the ``haze'' signal in the
  inner Galaxy, and argue that it does not follow a free--free
  spectrum as first thought, but instead is synchrotron emission from
  a hard electron cosmic-ray population.  3. In a departure from
  previous work, we allow the spectrum of \hal-correlated emission
  (which is used to trace the free--free component) to float in the
  fit, and find that it does not follow the expected free--free
  spectrum.  Instead there is a bump near 50 GHz, modifying the
  spectrum at the 20\% level, which we speculate is caused by spinning
  dust in the warm ionized medium.  4. The derived cross-correlation
  spectra are not sensitive to the map zero points, but are sensitive
  to the choice of CMB estimator.  In cases where the CMB estimator is
  derived by minimizing variance of a linear combination of the WMAP
  bands, we show that a bias proportional to the cross-correlation of
  each template and the true CMB is always present.  This bias can be
  larger than any of the foreground signals in some bands.  5.
  Lastly, we consider the frequency coverage and sensitivity of the
  \emph{Planck} mission, and suggest linear combination coefficients
  for the CMB template that will reduce both the statistical and
  systematic uncertainty in the synchrotron and haze spectra by more
  than an order of magnitude.
\end{abstract}

\keywords{ 
diffuse radiation ---
dust, extinction --- 
ISM: clouds --- 
radiation mechanisms: non-thermal --- 
radio continuum: ISM 
}

\section{Introduction}

Observations of the cosmic microwave background (CMB) by the
\emph{Wilkinson Microwave Anisotropy Probe} (WMAP) have revolutionized
our understanding of cosmology and placed strong constraints on
cosmological parameters \citep{spergel03,spergel07,tegmark04}.
Moreover, the WMAP foreground signal represents the most detailed and
sensitive full-sky maps of Galactic microwave emission, providing an
enormous wealth of information about the physical processes in the
interstellar medium (ISM).

\subsection{Galactic emission mechanisms}
There are three well established types of Galactic foreground 
signals at WMAP frequencies: free--free (or thermal bremsstrahlung) emission 
from interaction of free electrons with ions, dust emission from grains 
heated by the surrounding radiation field, and 
synchrotron emission from supernova shock accelerated electrons.  In addition, 
there are two other emission mechanisms which have proven more difficult to
characterize: spinning dust and the anomalous ``haze''
\citep{finkbeiner04}.  Spinning dust refers to emission from the
smallest dust grains which have non-negligible electric dipole moments
and are excited into rotational modes through a variety of mechanisms
\citep[cf.][]{DL98b}.  The physical origin of the haze is uncertain.

Though initially controversial \citep{bennett03}, numerous authors
have presented evidence for a spinning dust spectrum when combining
WMAP data with external data sets \citep{deO04,gb04,boughn07}.
Using only WMAP data, \citet{bennett03} and \citet{hinshaw07} point
out that it is difficult to spectrally distinguish certain spinning
dust models from synchrotron.  However, in our companion paper
\citep[][hereafter DF07]{DF07}, we show that a spinning dust spectrum
is indeed recoverable using exclusively WMAP data, though it is not
spatially correlated with the \emph{thermal} dust emission.

The haze was originally thought \citep{finkbeiner04} to be free--free emission 
from ionized gas which is too hot to be visible in recombination line maps and 
too cold to be visible in X-ray maps.  However, gas at the required temperature 
$T \sim 10^5$ K is thermally unstable \citep{spitzer}.  Furthermore, we will 
show in \refsec{results} that the spectrum of the haze is inconsistent with 
free--free emission and is most likely explained as a hard synchrotron component 
which is morphologically and spectrally distinct from the above mentioned 
\emph{soft} synchrotron.

\subsection{MEM analysis}
With the first and third year data releases, the WMAP team provided a dual 
foreground analysis: a maximum entropy method (MEM) and a template fitting 
algorithm.  The former was intended to improve our understanding of the 
astrophysics of foreground emission while the latter (more statistically stable) 
algorithm was meant to produce CMB maps with well characterized noise properties 
for use in the cosmological analysis \citep{bennett03,hinshaw07}.

Although the total observed emission matches the MEM model to better than 1\%, 
``Low residual solutions are highly constrained, but not necessarily unique or 
correct'' \citep[][p. 108]{bennett03}.  The MEM method is a pixel-by-pixel fit 
which minimizes the functional $H(p) = A(p) + \lambda(p)B(p)$ \citep{press92} 
where $\lambda$ is a regularizing parameter,
\be
  A(p) = \chi^2(p) = \sum_p[T(\nu,p) - T_m(\nu,p)]^2/\sigma^2,
\ee
\bel{mem}
  B(p) = \sum_c T_c(p) \ln[T_c(p)/P_c(p)],
\ee
and the sum is over Galactic emission components.  $P_c(p)$ is a prior template 
for component $c$, normalized to the same frequency as $T_c$ \citep[see][for 
details]{bennett03,hinshaw07}.

The noise properties of a MEM-derived CMB map are complicated, e.g. by the fact 
that noise is clipped to be non-negative by the logarithm in  \refeq{mem}.  
Because simple noise properties are desirable for a cosmological power spectrum 
analysis, a template-based map is used instead.  In the end, \citet{bennett03} 
and \citet{hinshaw07} fit spatial templates for the dust, free--free, and 
synchrotron ISM emissions to WMAP Q, V, and W bands.  They find that the 
remaining contamination in the maps is sufficiently subdominant outside the Kp2 
mask so that the effect on the CMB power spectrum is negligible.

\subsection{``Anomalous'' ISM emission mechanisms}
\label{sec:anomalous}

Given their dual approach, an important question to address is: why have the 
foreground analyses of the WMAP team not positively identified the spinning dust 
emission in DF07 and the haze emission in \citet{finkbeiner04}?

With the MEM analysis, there are two pitfalls.  First and most importantly, the 
MEM analysis generates a map of the spectral behavior of the soft synchrotron 
\emph{in each pixel}.  Since the free--free and dust spectra are kept fixed in 
the model, deviations away from these assumed prior spectra are absorbed into 
this ``spectral index map'' for synchrotron.  Second, when minimizing $H(p)$, if 
the recovered model $T_m(\nu,p)$ yields negative pixel values, $\lambda(p)$ is 
increased until the results are greater than zero.  These two effects 
combine so that foreground emission which does not match the prior templates and 
spectra (i.e., spinning dust and haze emission) are simultaneously absorbed 
into the synchrotron spectral index map and washed out by adding priors back 
into each pixel to enforce positivity.  Since minimizing $H(p)$ in this way is 
repeated multiple times (with the synchrotron spectral index map being 
``updated'' with each iteration) the results of the MEM analysis naturally 
strongly resemble the priors.

With the template fitting algorithm, the difficulty arises in the
choice of a synchrotron template.  Because the WMAP signal to noise
ratio is far superior to previous surveys, \citet{hinshaw07} use the
difference of the two lowest WMAP frequency maps (K$-$Ka) as a
template for synchrotron\footnote{
\citet{bennett03} used the synchrotron template described in 
\refsec{foremaps}, and an excess south of the Galactic Center is 
indeed visible in their residual maps (their Fig. 11, upper right panel).}.
They acknowledge that
this template necessarily contains free--free emission as well, but it
\emph{also} contains the haze and spinning dust emission.  Thus, the
haze and spinning dust are simultaneously explicitly fit with this
template.  In addition, the spectra of the dust emission and
free--free emission are again kept fixed so deviations are difficult
to identify.

Given the numerous challenges involved in a CMB foreground analysis and the 
serious questions about whether all the relevant foreground emission mechanisms 
have even been identified yet, we choose an approach that is simple enough to 
have well characterized noise properties, but flexible enough to allow us to 
find surprises.  In the limit where the spectrum of each component is invariant 
with position, one still has the choice of assuming a perfect spatial template, 
or assuming knowledge of the spectrum of each component.  Given that the 
spectrum of each component can vary with position, neither of these approaches 
is strictly correct.  Nevertheless, as we shall see, there is still much to be 
learned by making a too rigid assumption and then studying the resulting 
residuals.

\section{Methods}

We proceed under the assumption that the morphologies of the foreground 
components are well characterized by external data sets and that their spectra 
do not vary significantly over our regions of interest (see 
\refsec{regfits}).  Thus, in principle, we should be able to find the 
appropriate linear combination of the foreground templates which, when combined 
with a CMB template, can be subtracted from the WMAP data leaving only random 
residuals consistent with noise.

\subsection{Template maps}
\label{sec:foremaps}
Each of the templates used in our fits has been discussed in detail by previous 
authors \citep{bennett03,finkbeiner04,davies06,hinshaw07} and so we only 
briefly review them here.

\emph{Free--free:} Free--free (or thermal bremsstrahlung) emission originates 
from the Coulomb interaction of free electrons with ions in a warm gas.  Since 
this emission is proportional to gas density squared, maps of H$\alpha$ 
recombination line emission (which is also proportional to density squared) 
roughly trace the morphology of the gas and thus also the free--free emission.
Our template for this foreground is the H$\alpha$ map, assembled from the VTSS 
\citep{dennison98}, SHASSA \citep{gaustad01}, and WHAM \citep{haffner03} surveys 
by \citet{finkbeiner03}.  Extinction by dust (both in front of and
mixed with the warm gas) presents a potential challenge in
interpreting H$\alpha$ as a tracer of free--free emission
\citep{bennett03,finkbeiner04}.  We correct the \hal\ map for dust
extinction using the prescription in \citet{finkbeiner03}, and further
mitigate the effect by limiting our interest only to regions where the
dust extinction is $A(H\alpha) \equiv 2.65E(B-V) < 1$ mag, where $A$
is related to the dust optical depth $\tau_d = A/(1.086$ mag).  

The free--free spectrum is constrained by physics
\citep{spitzer,bennett03} as \be T \propto \nu^{\alpha},
  \label{eq:ffspec}
\ee
where $\alpha \sim -2.15$, $T$ is in antenna temperature, and the 
proportionality depends only on the electron temperature $T_e$.  Thus the only 
free parameter is a scaling factor equivalent to the electron temperature 
on the sky (but see \S \ref{sec:results}), though this temperature is expected 
to vary with position.

\emph{Thermal and spinning dust:} The emission produced from tiny 
interstellar dust grains 
vibrating in equilibrium with the surrounding radiation field has been mapped 
across the sky by \citet{schlegel98}.  We use their full-sky map evaluated at 94 
GHz by \citet{finkbeiner99} (hereafter, the FDS map) as a dust 
template.\footnote{the H$\alpha$ and FDS maps can be found online at 
http://www.skymaps.info}  The smallest of these dust grains are expected to have 
a non-negligible electric dipole moment and so can also emit radiation at WMAP 
frequencies through rotational modes excited by collisions with ions.  Thus our 
template also traces spinning dust emission.  Since the spectral dependence of 
the spinning dust is not well known, we use our fit to constrain the frequency 
dependence of the dust-correlated foregrounds.

\emph{Soft synchrotron:} As relativistic, shock accelerated electrons
travel through the Galactic magnetic field, they emit synchrotron
radiation with a characteristic frequency dependence $\propto
\nu^{\beta}$ (in antenna temperature units).  At 408 MHz, this
emission was measured by \citet{haslam82}, and we use their full-sky
map as a tracer of soft synchrotron.  As pointed out by
\citet{bennett03} and \citet{hinshaw07} the spectral index $\beta$ is
expected to vary across the sky, and in particular, the spectrum may
be harder near regions of recent supernova activity.  Though we use
our fit to evaluate the spectra of the 408 MHz-correlated synchrotron
emission, we note that a value of $\beta = -3.05$ removes most of the
emission at high latitude (most notably the prominent ``North Galactic
Spur'' feature).

\emph{CMB:} As we shall see in \refsec{results}, the choice of CMB
estimator can dramatically affect the inferred foreground spectra in a
given fit.  To illustrate this sensitivity, we use six different CMB
estimators defined as follows.
\begin{itemize}
\item
CMB1 -- The published internal linear combination (ILC) map derived by the WMAP 
team for the 3-yr data.\footnote{available at \texttt{http://lambda.gsfc.nasa.gov/}}
\item CMB2 -- An ILC with the coefficients that the WMAP team have
  found best cancel their Region 0 foregrounds in the three year data
  \citet{hinshaw07}.  This ILC is given by,
\be
  \begin{array}{ccl}
  {\rm CMB2} & = & 0.156 \ T_{\rm K} \ - \ 0.888 \ T_{\rm Ka} \ + \ 0.030 \ 
                   T_{\rm Q} \\
    &   & + \ 2.045 \ T_{\rm V} \ - \ 0.342 \ T_{\rm W},
  \end{array}
\ee
where $T_j$ is the observed WMAP temperature data in band $j$, in thermodynamic $\Delta T$ units. 
\item
CMB3 -- An ILC which minimizes the variance over our unmasked pixels.  Here
\be
  \begin{array}{ccl}
  {\rm CMB3} & = & -0.032 \ T_{\rm K} \ - \ 0.205 \ T_{\rm Ka} \ + \ 0.037 \ 
                   T_{\rm Q} \\
    &   & + \ 0.441 \ T_{\rm V} \ + \ 0.760 \ T_{\rm W}.
  \end{array}
\ee
\item
CMB4 -- A ``high frequency (HF) estimator'' that removes the dominant 
foregrounds (thermal dust and free--free) from the 94 GHz WMAP data,
\be
  \begin{array}{ccl}
  {\rm CMB4} & = & T_{\rm W} - FDS - A H\alpha,
  \end{array}
\ee
where the constant $A$ is determined from the approximate free--free amplitude at 
23 GHz and traced to 94 GHz via \refeq{ffspec}.
\item
CMB5 -- A model for the thermal dust is subtracted from all of the WMAP bands 
using $T_{\rm dust} = (\nu/94\mbox{ GHz})^{1.7} \times$ FDS.  A minimum variance 
ILC is then generated from this thermal dust pre-subtracted data (denoted by 
primes).  The variance is minimized for,
\be
  \begin{array}{ccl}
  {\rm CMB5} & = & 0.104 \ T'_{\rm K} \ - \ 0.289 \ T'_{\rm Ka} \ 
                  - \ 0.190 \ T'_{\rm Q} \\
    &   & + \ 0.317 \ T'_{\rm V} \ + \ 1.059 \ T'_{\rm W},
  \end{array}
\ee
\item
CMB6 -- A cleaned map of the 3-year data, cleaned with the \citet{tegmark03} 
(TOH) method.\footnote{available at http://space.mit.edu/home/tegmark/wmap.html}  
This method utilizes a linear weighting of the data in which the weights depend 
on the multipole $\ell$ of the spherical harmonic expansion of each of the five 
WMAP bands.
\end{itemize}

In the limit where the noise in each WMAP band is equal to 
$\sigma_0$, the measurement noise of the ILC is simply
\be
\sigma_L = \sigma_0\sqrt{\sum_b \zeta_b^2}
\ee
so CMB3 and CMB5 are significantly less noisy than CMB1 and CMB2.  Due to the 
complicated weighting of the TOH method, the measurement noise properties 
of CMB6 are quite complicated.

\emph{Mask:} In addition to masking out all point sources listed in
the WMAP team's three year catalog, as noted above, we mask all
regions of the sky where the H$\alpha$ extinction due to dust
$A(\mbox{H}\alpha)=2.65 E(B-V) \ge 1$ mag. We also mask out the LMC,
SMC, M31, Orion-Barnard's Loop, NGC 5090, and $\zeta$--Oph.  This mask
covers $21.5\%$ of the sky.

\subsection{Fitting procedure}
\label{sec:fitpro}
Our model is that the observed WMAP data is a linear combination 
of the foreground templates plus the CMB\footnote{Variations in the foreground 
spectra from place to place on the sky will be explored by fitting smaller 
regions in upcoming sections} plus noise.  Therefore, we want to solve the 
matrix equation 
\bel{mateq}
  P \rmbf{a} = \rmbf{w},
\ee
where \rmbf{w} is the CMB-subtracted WMAP data and $P$ is a ``template 
matrix'' whose columns 
consist of the foreground templates outlined in \S \ref{sec:foremaps}, for the 
coefficient vector \rmbf{a} whose entries represent the weights of the 
individual foregrounds.

The template matrix can be represented schematically by a block-diagonal matrix,
\be
P=
\left(
  \begin{array}{ccccc}
    P_1 & & & & \\
    & P_2 & & & \\
    & & P_3 & & \\
    & & & P_4 & \\
    & & & & P_5
  \end{array}
  \right),
\ee
where the 5 blocks correspond to the 5 WMAP bands, and each block has
the form
\be
  P_b=
  \left(
  \begin{array}{ccc}
  f_{1,b} & d_{1,b} & s_{1,b} \\
  f_{2,b} & d_{2,b} & s_{2,b} \\
  f_{3,b} & d_{3,b} & s_{3,b} \\
  . & . & . \\
  . & . & . \\
  . & . & . \\
  f_{N_p,b} & d_{N_p,b} & s_{N_p,b}
  \end{array}
  \right),
\ee
where \rmbf{f}, \rmbf{d}, and \rmbf{s} are the templates for free--free, 
thermal and spinning dust, and soft synchrotron emission (in thermodynamic mK)
respectively.  This makes $P$ a $5 N_p \times 15$ matrix.  For each
template, the mean of the unmasked pixels is subtracted, making the
results of this fit insensitive to zero-point errors in the templates.
For each template column, the first index represents the pixel number
and the second represents a WMAP frequency band --- i.e., 1=23 GHz
(K), 2=33 GHz (Ka), etc.  The total number of un-masked pixels in each
map is
$N_p$.  For our most general fits, we assume no knowledge of the \rmbf{f}, 
\rmbf{d}, and \rmbf{s} 
spectra, and so those templates do not differ for each band.  Rather, we 
explicitly \emph{fit} the spectra as discussed below.

The CMB-subtracted WMAP data and the coefficient vector are column vectors,
\be
  \rmbf{w} = 
  \left(
  \begin{array}{c}
  T_{1,1} - c_1 \\
  T_{2,1} - c_2 \\
  T_{3,1} - c_3 \\
  . \\
  . \\
  . \\
  T_{1,2} - c_1 \\
  . \\
  . \\
  . \\
  T_{N_p,5} - c_{N_p}
  \end{array}
  \right) 
  \mbox{ \ \ and \ \ }
  \rmbf{a} = 
  \left(
  \begin{array}{c}
  a_{f,1} \\
  a_{d,1} \\
  a_{s,1} \\
  a_{f,2} \\
  a_{d,2} \\
  a_{s,2} \\
  . \\
  . \\
  . \\
  a_{s,5} \\
  \end{array}
  \right),
\ee
where $T_{i,j}$ is the observed WMAP temperature data in pixel $i$ and band $j$, 
and \rmbf{c} is one of our CMB estimators.  Their lengths are $5 N_p$ and 15 
respectively.

Since $P$ is not a square matrix with linearly independent rows, it is not 
invertible.  To solve \refeq{mateq} we calculate $P^+$, where $^+$ denotes the 
\emph{pseudoinverse}.\footnote{The pseudoinverse is defined as, $P^+ = B 
\Sigma^+ U^T$, where the singular value decomposition of $P = U \Sigma B^T$ and 
$\Sigma^+$ is the transpose of $\Sigma$ with all non-zero singular values 
replaced by their inverse.  In the case of a square, non-singular matrix, $P^+ 
\rightarrow P^{-1}$.}  The solution $\rmbf{a} = P^+ \rmbf{w}$ minimizes the 
quantity $e^2 = ||P\rmbf{a}-\rmbf{w}||^2$, so that if we divide both sides 
of \refeq{mateq} by the uncertainty 
$\sigma$ (following Bennett et al. 2003 and Hinshaw et al. 2007, 
we use the mean measurement noise in each WMAP band), the solution\footnote{From 
the properties of the pseudoinverse, 
$P P^+ P = P$ and $[P P^+]^T = P P^+$, it is easy to show that, if the columns 
of $P$ are linearly independent as they are in our case, $P^+ = [P^T P]^{-1} 
P^T$, making our technique equivalent to other $\chi^2$ minimization techniques 
\citep[e.g.][]{tegmark03,deO06}.
}
\bel{fitsol}
  \rmbf{a} = \left( P/\sigma\right)^+ \left( \rmbf{w}/\sigma \right)
\ee
minimizes the quantity
\bel{chisq}
  \left\|\frac{P}{\sigma} \ \rmbf{a} - \frac{\rmbf{w}}{\sigma}\right\|^2 = 
  \frac{\|P\rmbf{a}-\rmbf{w}\|^2}{\sigma^2} \equiv \chi^2.
\ee

This fitting procedure is flexible in that additional foreground components can 
be incorporated by simply adding columns to the template matrix.  We will 
exploit this feature in \refsec{results} to model the anomalous ``haze'' excess 
emission towards the Galactic center.

The five components of each coefficient (e.g., $\rmbf{a}_{d,j}$, with j=[1:5]) 
represent a fit of both the amplitude and spectrum of 
the associated emission.  By simultaneously fitting all of the spectra for all 
of the foregrounds, we can completely decouple the bands from each other in our 
fits.  Additionally, we also perform less general fits in which various spectra 
are fixed to follow the dependencies in \refsec{foremaps} (see 
\reftbl{fittypes}).  We point out that, fitting the spectrum of the free--free 
emission serves as a check on the assumption that it is well described by 
Equation \ref{eq:ffspec} (see \refsec{results}).

\section{Results}
\label{sec:results}
We have performed both full-sky fits as well as fits of smaller regions --- the 
motivation being that both the soft synchrotron and spinning dust spectra should 
vary from place to place across the sky.  Our fits are characterized by the 
$\chi^2$ statistic of \refeq{chisq} and by the residual map,
\be
  \Res = P\rmbf{a} - \rmbf{w},
\ee
which guides intuition and serves as a visual aid in evaluating the goodness of 
fit.

\begin{deluxetable}{c|cccc|ccc}
\tablehead{
  & \multicolumn{4}{c}{Spectra Fit} & \multicolumn{3}{|c}{$\chi^2/\nu$} \\
  Type & H$\alpha$ & Dust & Haslam & Haze$^{\dagger}$ & FS & GC & 
  RG$^{\ddagger}$
}
\startdata
  1 &   & x &   &   & 3.514 & 6.700 & 4.572 \\
  2 &   & x &   & x & 2.993 & 5.147 & 4.213 \\
  3 & x & x &   &   & 3.498 & 6.656 & 4.302 \\
  4 & x & x &   & x & 2.977 & 5.126 & 4.168 \\
  5 &   & x & x &   & 3.506 & 6.650 & 4.241 \\
  6 &   & x & x & x & 2.988 & 5.106 & 4.169 \\
  7 & x & x & x &   & 3.489 & 6.611 & 4.208 \\
  8 & x & x & x & x & 2.972 & 5.082 & 4.148
\enddata
\tablecomments{
The different types of fits performed with the procedure outlined in 
\refsec{fitpro} using CMB5 for a CMB estimator.  Fit types are characterized by 
which spectra are fit, and in which region of the sky: full-sky (FS), Galactic 
center (GC, $l,b=[-45:+45]$), or specific regions of interest (RG, see 
\reffig{regions-map}).  $\dagger$ The haze template is described in 
\refsec{fs-and-gc-fits}.  $\ddagger$ The $\chi^2/\nu$ for the RG fits are 
averaged over all 12 regions.
}\label{tbl:fittypes}
\end{deluxetable}

The fits can be separated into eight types as outlined in \reftbl{fittypes}.  
The types designate which spectra are fit and whether the fit was over 
the full-sky (FS), the Galactic center (GC -- $l,b=[-45:+45]$), or the 
``regions'' (RG), which were chosen as regions of interest based on $\Res$ for a 
full-sky fit (see \reffig{regions-map}).  Also shown in \reftbl{fittypes} are 
the $\chi^2/\nu$ for each fit type, where $\nu$ is the number of degrees of 
freedom and is given by
\be
  \nu = N_p - N_a,
\ee
with $N_a$ the number of elements in the coefficient vector.  Throughout the rest 
of this paper, we concentrate exclusively on our most general fit types 7 and 8, 
and unless otherwise noted, use CMB5 for a CMB estimator.  This estimator is 
attractive both because of its uncomplicated noise properties and because the 
thermal dust emission has arguably the most well understood amplitude and 
spectral behavior making it amenable to pre-subtraction before forming an ILC.

\subsection{Full-sky and Galactic center residual maps}
\label{sec:fs-and-gc-fits}

\bpm
\centerline{
  \includegraphics[width=0.9\textwidth]{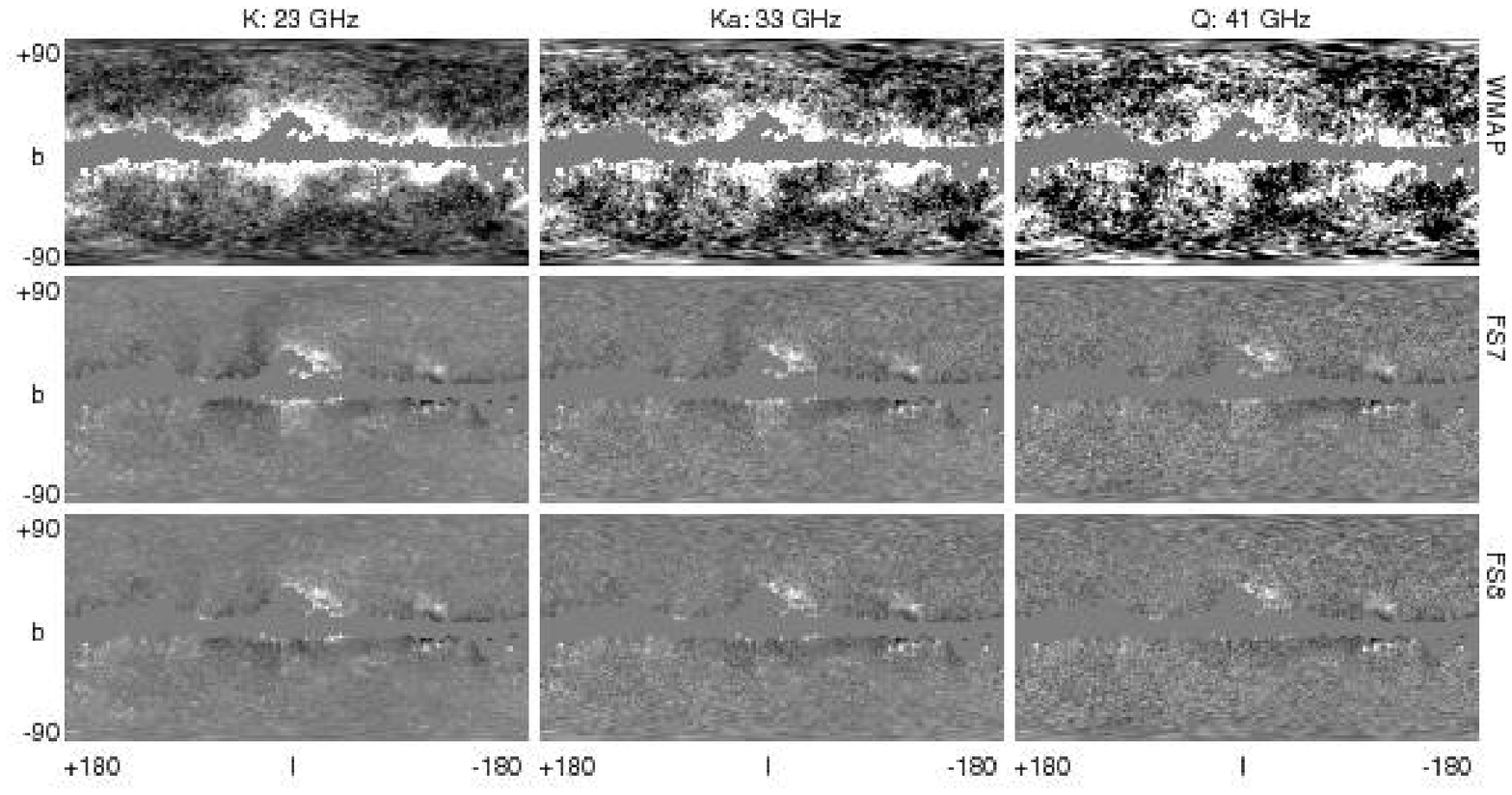}
}
\caption{
Residual maps $\Res$ for three of the five bands and for FS7 and FS8 fits using 
CMB5 (the K, Ka, and Q  band maps are stretched to $\pm 0.25$, $\pm 0.12$, and 
$\pm 0.08$ mK respectively; all maps are mean subtracted).  
The unsubtracted WMAP data are also shown for comparison.  The FS7 fit removes 
much of the 
emission, however there is a remaining excess towards the Galactic center.  This 
excess emission is particularly notable south of the Galactic center where 
obscuration by dust and gas is negligible.  FS8 incorporates a simple spatial 
template for this haze and removes much of the residuals in that region.
}\label{fig:fullsky}
\epm

\reffig{fullsky} shows $\Res_{\rm FS7}$ as well as the unsubtracted
maps for K, Ka, and Q bands, stretched to $\pm 0.25$, $\pm 0.12$, and $\pm 0.08$ 
mK respectively.  The fit yields
$\chi_{\rm FS7}^2/\nu = 3.49$, removing 95.8\%, 95.7\%, 96.3\%,
97.5\%, and 99.7\% (for K, Ka, Q, V, and W\footnote{The noise in W
band is especially
low because the coefficient of $T_W$ in the CMB5 estimator is very
close to one, so the CMB-subtracted W band data has essentially no
W band data in it.  Therefore results from W band should be viewed
with suspicion.} bands respectively) of the
variance from the WMAP data.  However, it is clear from the second row
of \reffig{fullsky} that there is still a remaining emission residual
towards the Galactic center (GC).  
This residual is the 
``haze'' present in the 1 year data as shown by \citet{finkbeiner04}.

Although the average power in the haze is small (with a mean of just 0.59 kJy/sr 
per pixel at 23 GHz within 30 degrees of the GC in the southern sky), our fit may 
be compensating for its presence by adjusting the weights of the other 
foreground templates.  To relax the stress on the fit, we adopt a crude model for 
the haze emission,
\bel{haze-mod}
  \rmbf{h} \propto \left\{
  \begin{array}{cl}
  \frac{1}{r} - \frac{1}{r_0} & \mbox{for } r < r_0; \\
  0 & \mbox{for } r > r_0,
  \end{array}
  \right.
\ee
where $r$ is the distance to the Galactic center and we arbitrarily
set $r_0 = 45$ degrees.  Since the emission mechanism is unknown, we
fit the spectrum of the haze as well.  The extent to which the other
fit parameters change gives an idea of the cross talk between the haze
and the other templates.

The third row of Figure \ref{fig:fullsky} shows residual maps for FS8.  It is 
clear from $\Res_{\rm FS8}$ alone that the fit is improved, particularly in the 
southern GC where obscuration from dust and gas is minimal and in the high 
latitude north where the North Galactic Spur synchrotron feature is no longer over 
subtracted at 23 GHz (compare rows 2 and 3 of Figure \ref{fig:fullsky}).  Though 
$\chi^2/\nu_{\rm FS8} = 2.97$ is only slightly lower than the FS7 fit, we 
emphasize that the number of degrees of freedom is quite large ($\nu \sim 
155,000$) and so the likelihood for the FS8 model is significantly higher.  
Furthermore, we are including many pixels at large Galactic latitudes where the 
signal to noise is very low and the amplitude of the haze is very small.

\bp
  \includegraphics[width=0.45\textwidth]{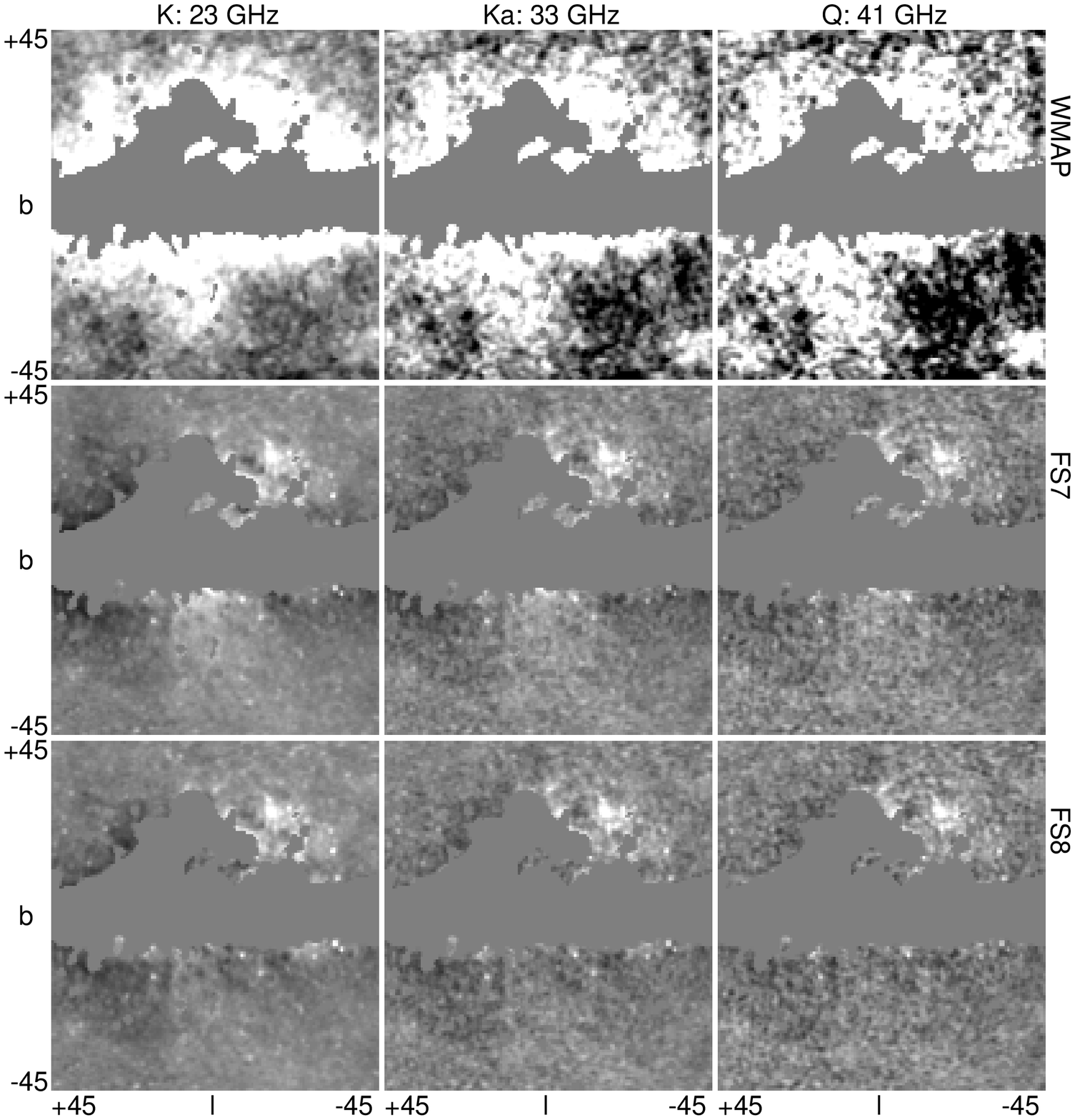}
\caption{
The same as \reffig{fullsky} except for the GC fits ($l,b = [-45:45]$ degrees).  
Though GC7 explicitly fits the free--free, dust, \emph{and} synchrotron spectra, 
the haze is still present in $\Res_{\rm GC7}$ indicating that it is 
morphologically dissimilar to the those templates.  The quality of the fit is 
substantially improved with the inclusion of a haze template (GC8).  The extended 
structure in the north-west GC is likely due to an imperfect haze template.
}\label{fig:gc}
\ep

Since the ISM may have different properties near the GC (i.e., lower gas 
temperature due to more efficient cooling because of higher metallicity, 
increased supernova activity, etc.), it is instructive to consider this region 
separately from the rest of the sky.  \reffig{gc} shows $\Res$ maps for 
the GC7 and GC8 fits.  There is a significant decrease in $\chi^2/\nu$ from GC7 
to GC8 with the inclusion of our haze template --- from 6.61 for GC7 to 5.08 for 
GC8.  The substantially oversubtracted regions near the edge of the mask in 
$\Res_{\rm GC7}$ indicate that the fit is indeed attempting to compensate for 
the haze by adjusting the amplitudes of other templates.  Although there is 
still some over subtraction in $\Res_{\rm GC8}$, the overall quality of the fit 
is much improved, particularly in the southern sky.

In the \emph{northern} sky, there is a large structure just north-west of the GC 
in and around the region of Rho-Oph.   Typically, extended regions of over- or 
under-subtraction in our fits are indicative of variations in the physical 
conditions of the emission media.  However, in this circumstance, it is unclear 
if the under-subtraction is due to an over simplification of the radial haze 
profile given our template for this component (\refeq{haze-mod}).  The next step 
is to subdivide the sky into certain regions of interest where our fit is the 
least successful and fit those regions explicitly.

\subsection{Regional fits and composite maps}
\label{sec:regfits}
\bp
  \includegraphics[width=0.45\textwidth]{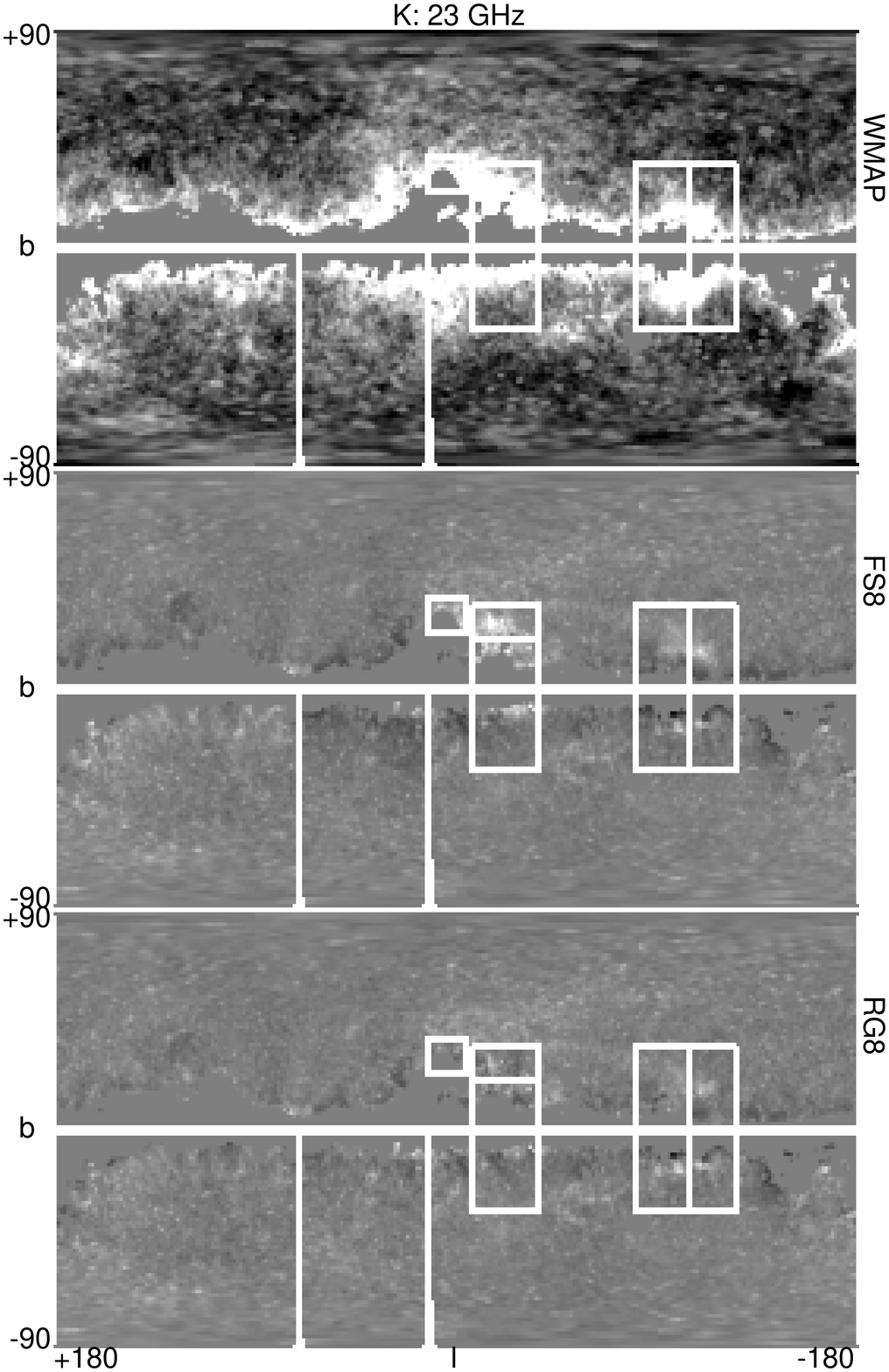}
\caption{
Illustration of the 12 regions independently fit in the RG fits.  Breaking the sky 
up in this way removes many of the over- and under-subtracted features in the 
residual maps, particularly the excess emission in the north-west GC.
}\label{fig:regions-map}
\ep

\bpm
\centerline{
  \includegraphics[width=0.9\textwidth]{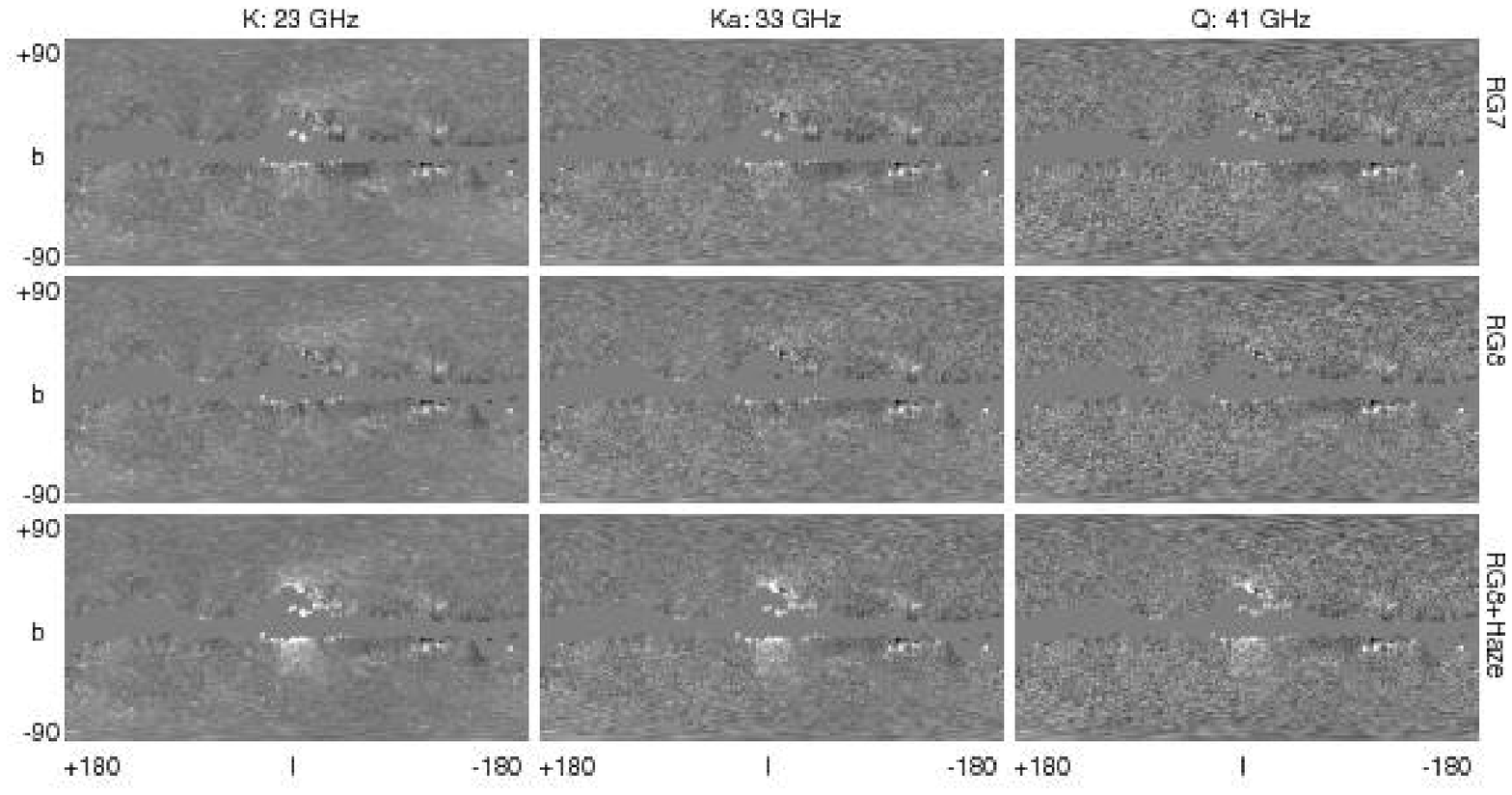}
}
\caption{
Residual maps for the RG7 and RG8 fits using CMB5, with the same stretch as 
\reffig{fullsky}.  The haze is still present in the $\Res_{\rm RG7}$ fits 
despite the relatively small size of the fitting regions.  The bottom row shows 
$\Res_{\rm RG8}$ with the subtracted haze in each region added back in to that 
region.  These maps are unsmoothed and no continuity constraints are placed on 
the region boundaries.
}\label{fig:fullsky-regions}
\epm

Our regions of interest were identified as regions of particularly notable over- or 
under-subtraction in $\Res_{FS8}$.  \reffig{regions-map} shows the boundaries of 
these regions superimposed on the raw WMAP data at 23 GHz, $\Res_{\rm FS8}$, as 
well as a composite map for $\Res_{\rm RG8}$ (a fit which includes our haze 
template).  The residual map (which is actually the residual maps of each region 
stitched together with \emph{no} smoothing) is shown for both RG8 and RG7 fits 
in \reffig{fullsky-regions} (the maps stretched to the same units as 
\reffig{fullsky}).

At all frequencies, the RG8 fit more effectively removes the foregrounds than 
the RG7 fit.  There are more regions of over-subtraction in $\Res_{\rm RG7}$ 
(both at high latitudes and around the mask edges) compared to $\Res_{\rm RG8}$, 
the region boundaries are somewhat less continuous in $\Res_{\rm RG7}$, and most 
importantly, the haze is \emph{still} present in the $\Res_{\rm RG7}$ maps.  
These composite $\Res$ maps are completely unsmoothed and it is a testament to 
the quality of our RG8 fit that there are no discernible large scale brightness 
gradients between adjacent regions.  

Lastly, we define the residual haze map
\be
  \Res_{H} = \Res_{\rm RG8} + \rmbf{a}_{h} \rmbf{h},  
\ee
where the appropriate amount of haze is added back in to each region (i.e., the 
same amount as was subtracted by the fit).  The bottom row of 
\reffig{fullsky-regions} shows $\Res_H$.  Despite the fact that the haze fit 
coefficients are not constrained to be continuous across the region boundaries, 
$\Res_H$ has no clear discontinuities.

\subsection{Foreground spectra}

\bpm
\centerline{
  \includegraphics[width=0.45\textwidth]{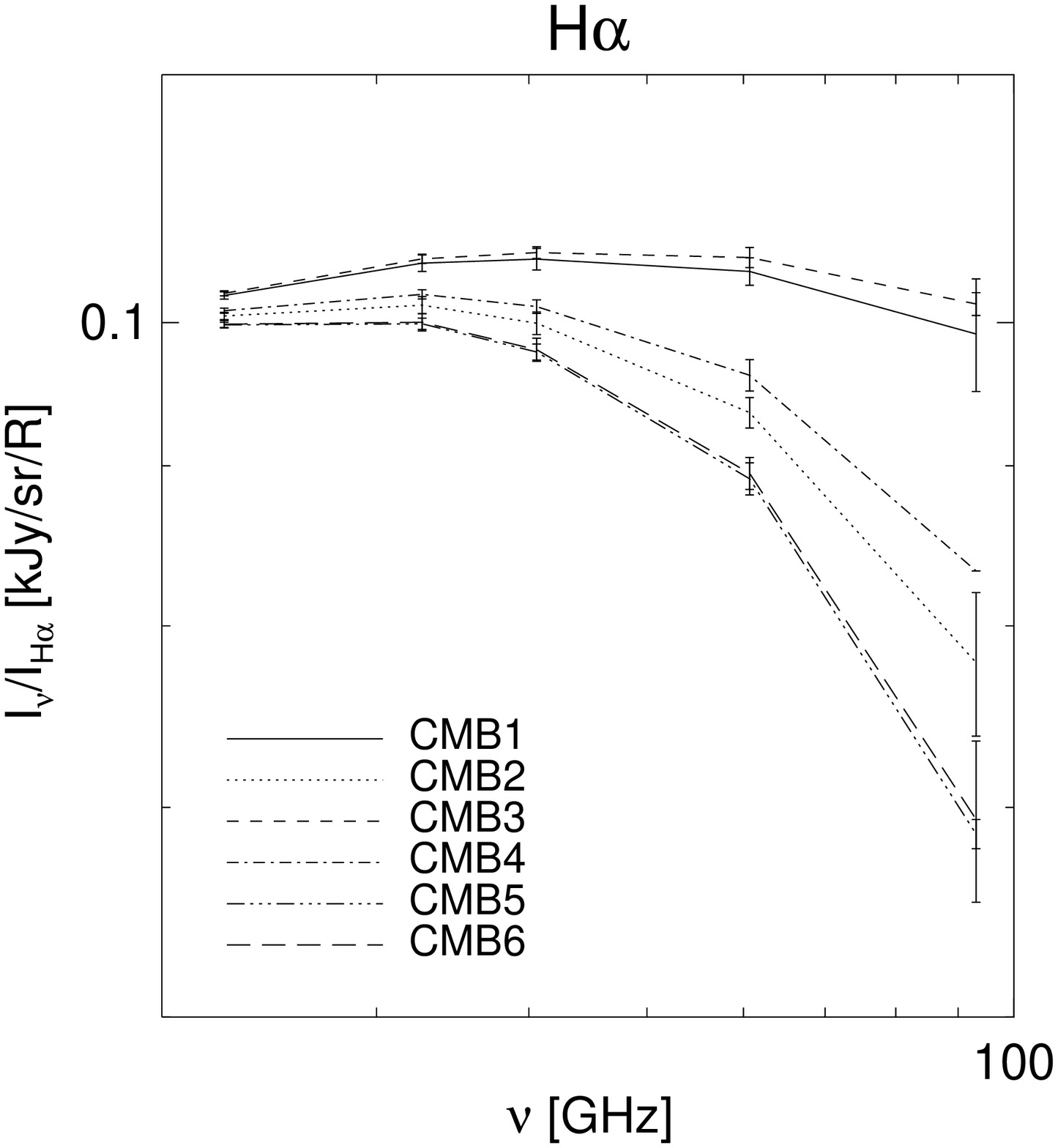}
  \includegraphics[width=0.45\textwidth]{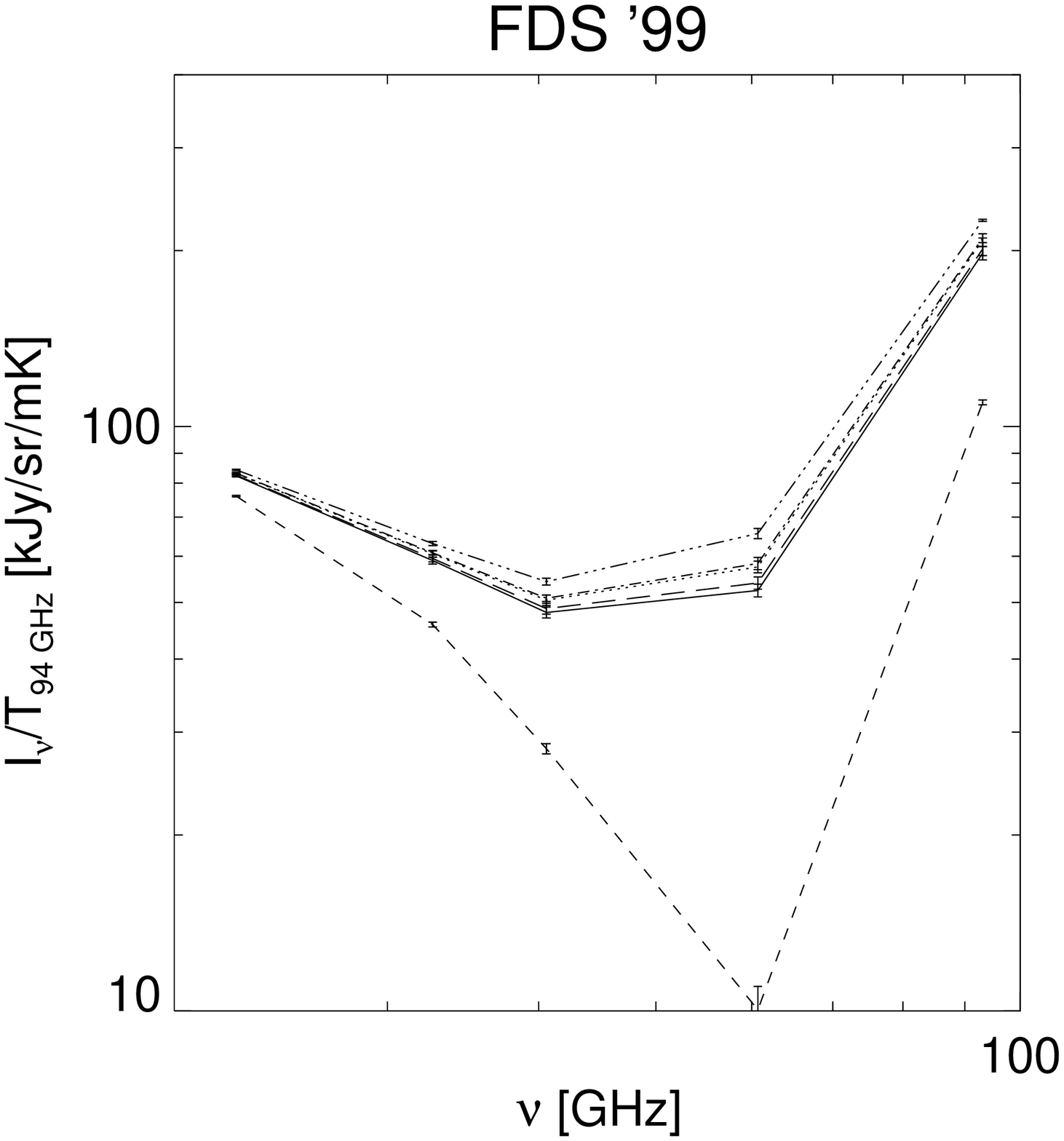}
}
\centerline{
  \includegraphics[width=0.45\textwidth]{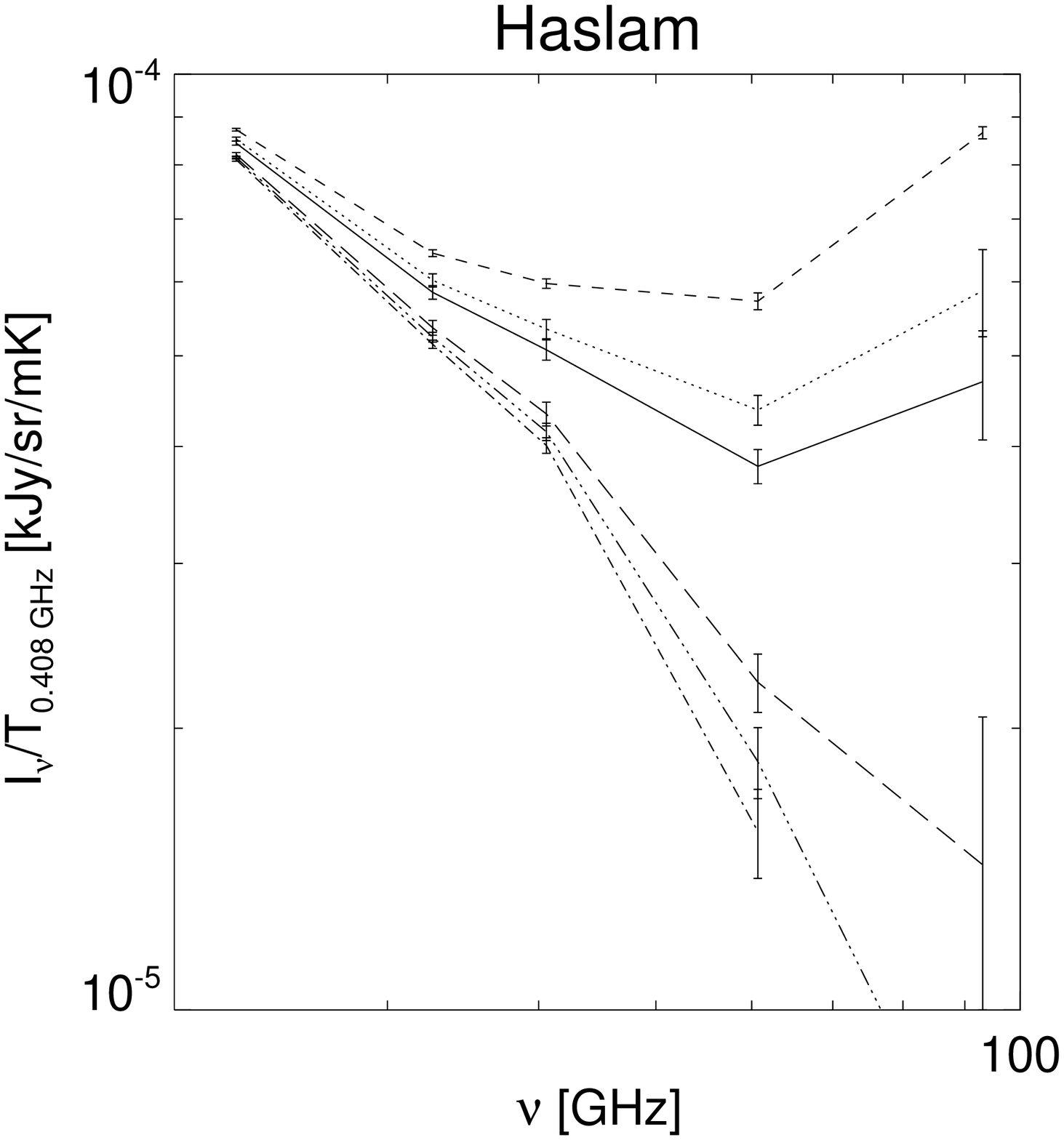}
  \includegraphics[width=0.45\textwidth]{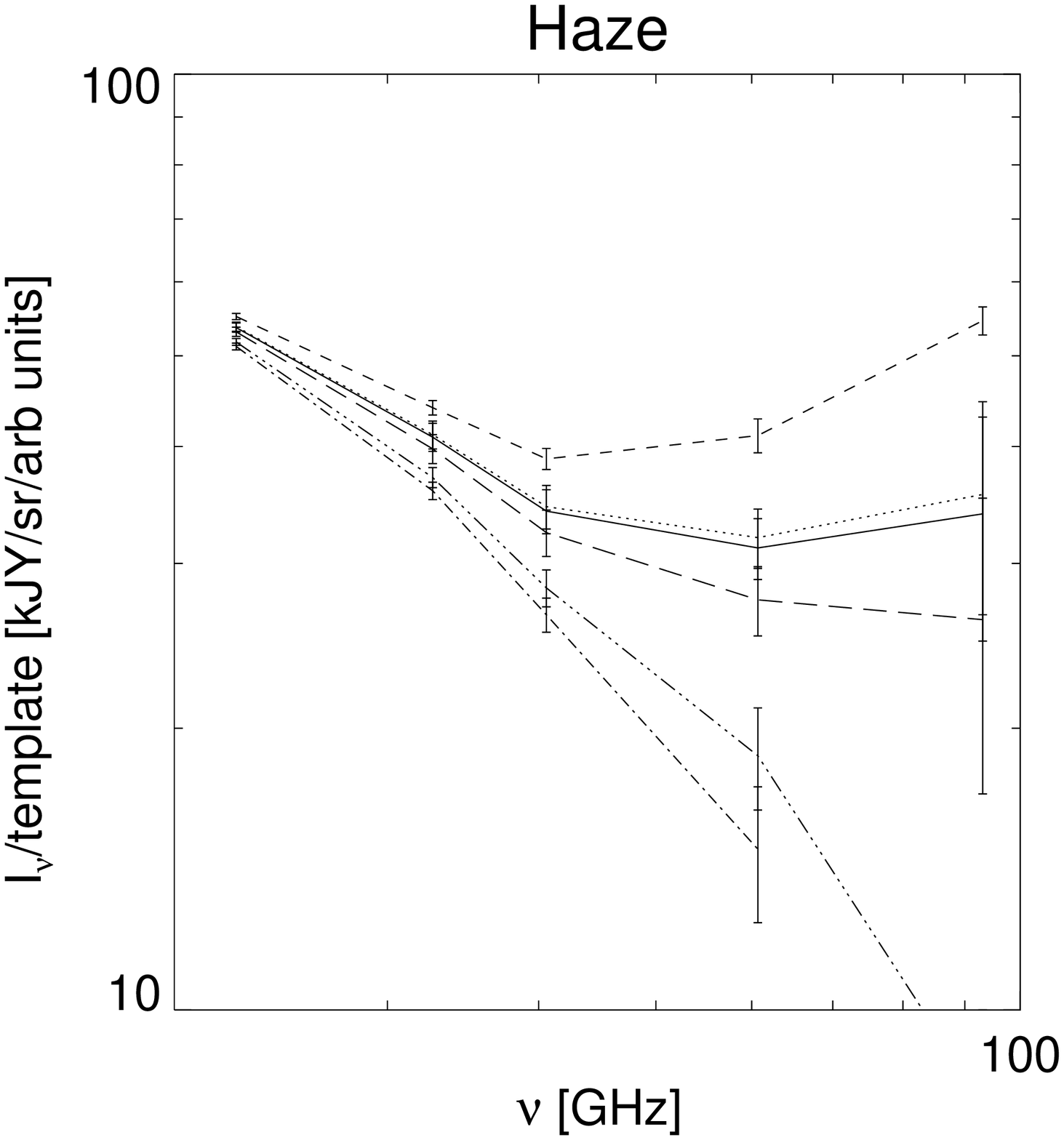}
}
\caption{
Foreground spectra for the FS8 fits with our six CMB estimators: solid = CMB1, 
dotted = CMB2, short dashed = CMB3, dot-dashed = CMB4, dot-dot-dot-dashed = 
CMB5, and long dashed = CMB6.  The error bars on the coefficients are the formal 
error bars on the fit (see text).
}\label{fig:spectra-fs}
\epm

\reffig{spectra-fs} shows the FS8 spectra (coefficient vectors) in kJy/sr per 
template unit: Rayleighs for H$\alpha$, mK for FDS and Haslam, and arbitrary 
units for the haze.  The most striking feature of these spectra are that they 
are very sensitively dependent on the estimator used for the CMB.  For example, 
the synchrotron spectrum actually appears to turn \emph{up} at high frequencies 
for CMB1, CMB2, and CMB3, while CMB4 and CMB5 give the more physically motivated 
power law type spectra.  This dependence is entirely due to the contamination of 
the CMB by foregrounds, which can never fully be removed for any CMB estimator.  
Thus, when we remove the estimator from the WMAP data to perform the foreground 
fit, we have inadvertently added (or subtracted) some foregrounds from the data 
with essentially the spectrum of the CMB ($I_{\nu} \propto \nu^{2}$).  We 
emphasize that the contamination is small, so that it has minimal effect on the 
\emph{variance} of the CMB estimator, but it is large compared to the relative 
amplitudes of the foregrounds.  This is especially true for estimators which 
minimize the variance of an ILC; in this case the contamination is proportional to 
the cross correlation of the true CMB with the true foregrounds 
(see \refsec{ilc-bias}).

Despite the large uncertainties, there are concrete conclusions that can be drawn 
from \reffig{spectra-fs}.  First, the H$\alpha$-correlated emission does 
\emph{not} follow the $I_{\nu} \propto \nu^{-0.15}$ frequency dependence as 
expected.  Instead there is a bump in the spectrum around 30 GHz.  In our 
companion paper \citep{DF07} we argue that the H$\alpha$-correlated 
emission has a spectrum that is consistent with a classical $\nu^{-0.15}$ 
spectrum plus a WIM spinning dust component.  Second, although the soft 
synchrotron and haze spectra vary substantially with CMB estimator type, for a 
given type, the haze is always \emph{harder} than the normal soft synchrotron, a 
point which we explore in more detail below.

Finally, like \citet{bennett03}, \citet{finkbeiner04}, \citet{davies06}, and 
\citet{hinshaw07}, we find that the dust-correlated emission falls from 94 to 61 
GHz but then rises to 23 GHz consistent with emission from both thermal and 
spinning dust.  Since these spectra are the result of fits over large areas of 
the sky and the spinning dust spectrum is expected to vary with position, it is 
not surprising that we do not see a peak in the dust-correlated emission in the 
range 20-40 GHz as in the \citet{DL98b} models.  Rather, we are seeing a 
superposition of many spinning dust spectra with varying peak frequencies.  This 
has led to the misidentification of the this dust-correlated emission as 
synchrotron in the past \citep{bennett03,hinshaw07}.

It is tempting to conclude that the RG fits for the individual regions can be 
used to construct a map of the variation in the spinning dust spectra across the 
sky.  Such a map would represent an ``excitation map'' or ``irradiation map'' 
for the dust grains.  However, this is simply not possible given the 
contamination of the CMB estimator by the foregrounds.\footnote{Uncertain zero 
point variation across the maps (due to instrumental limitations, imperfect 
zodiacal light subtraction, etc.) is also a consideration, but it is 
subdominant.}

\subsection{Comparison of the haze and soft synchrotron}
\label{sec:haze-vs-sync}

From \reffig{spectra-fs} it seems that the haze spectrum is inconsistent with a 
$\nu^{-0.15}$ free--free spectrum.  While some of the CMB estimators do yield 
such hard spectra at low frequencies, those fits turn up at high frequencies.  
This behavior is \emph{much} less likely than simply a contamination of the CMB 
estimator by the haze emission.  Furthermore, the haze is most clearly visible 
in the southern GC where dust obscuration is minimal so that the haze should 
either show up in the H$\alpha$ map (if it is $T < 10^5$ K gas) or in the ROSAT 
X-ray data (if it is $T > 10^6$ K gas).  Since neither of these is the case, and 
since gas at intermediate temperatures is thermally unstable 
\citep[see][]{spitzer}, we concluded that the most likely source of the haze is 
synchrotron emission from relativistic electrons.  \reffig{spectra-fs} also 
shows that this synchrotron emission is harder than the soft synchrotron traced 
by the Haslam 408 MHz map.\footnote{In their first and third year temperature 
analysis, the WMAP team concludes that the spectrum of the \emph{soft} 
synchrotron is harder near the Galactic plane and particularly near the Galactic 
center \citep{bennett03,hinshaw07}.  This may be inconsistent with 
\citet{page07} and \citet{kogut07} who find a more or less constant synchrotron 
spectral index across the sky from the WMAP polarization data.  Given the 
difficulties in the interpretation of the MEM analysis (see \refsec{anomalous}), 
the reason for the discrepancy is unclear.}

\bpm
\centerline{
  \includegraphics[width=0.3\textwidth]{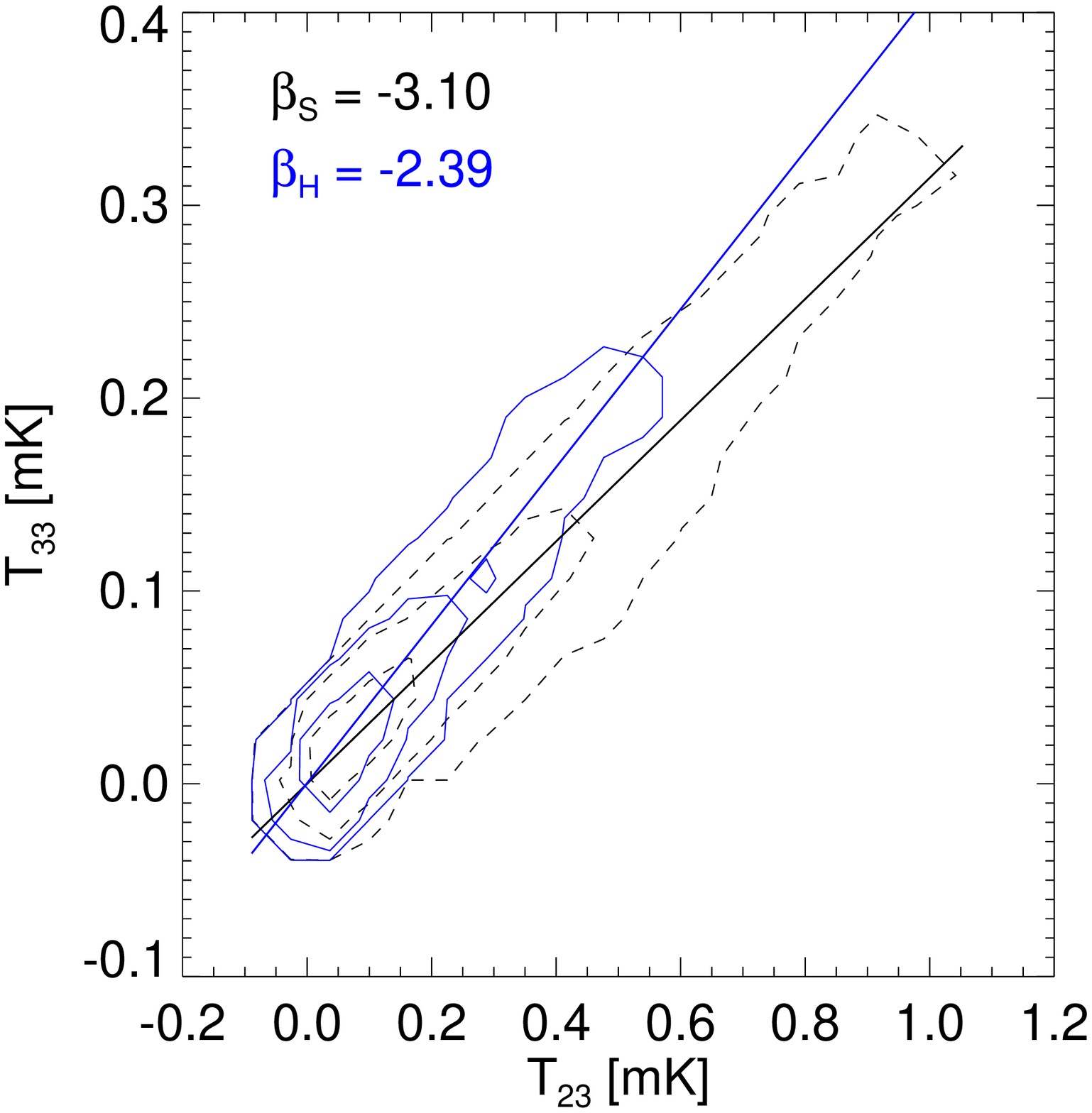}
  \includegraphics[width=0.3\textwidth]{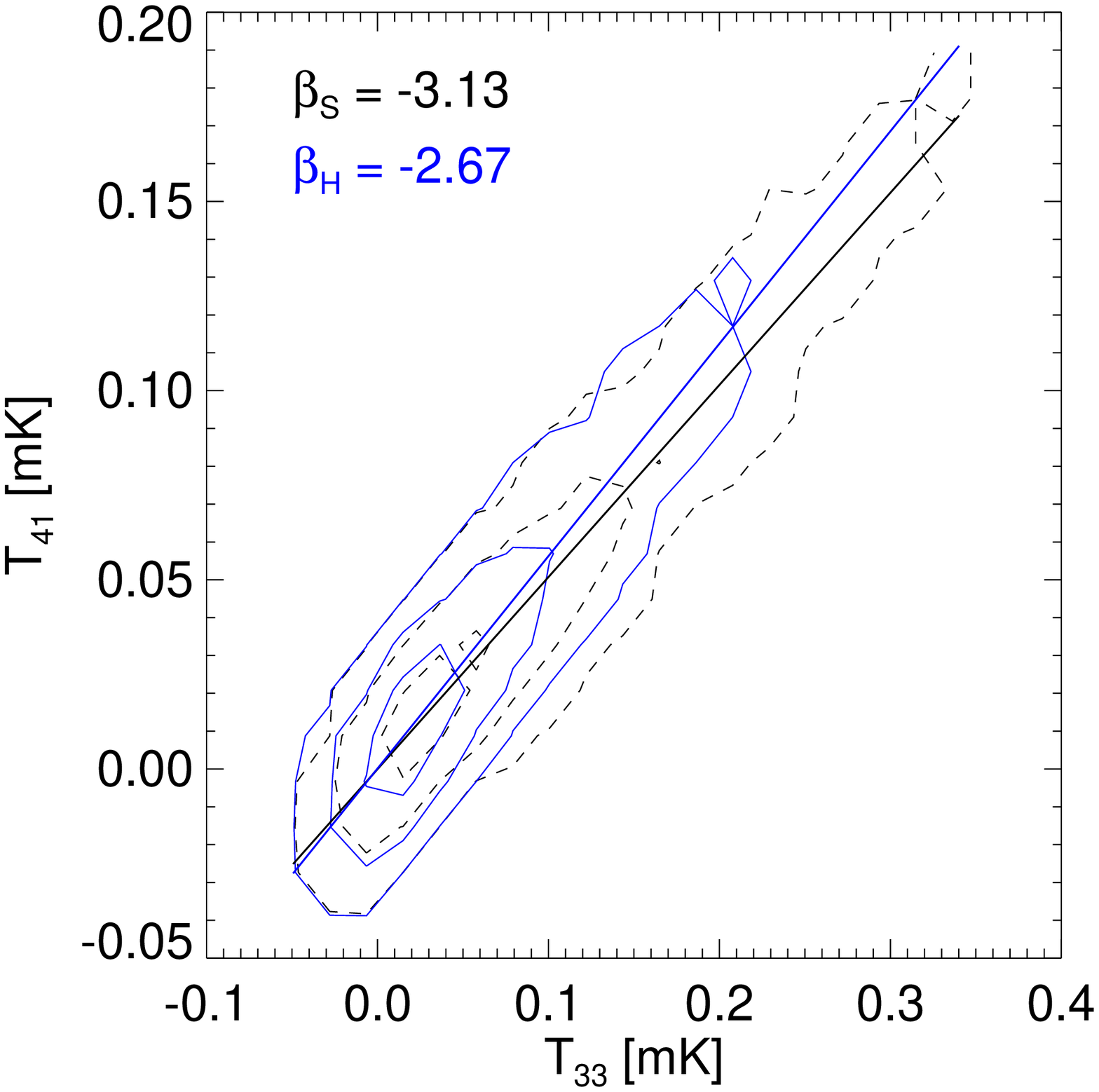}
  \includegraphics[width=0.3\textwidth]{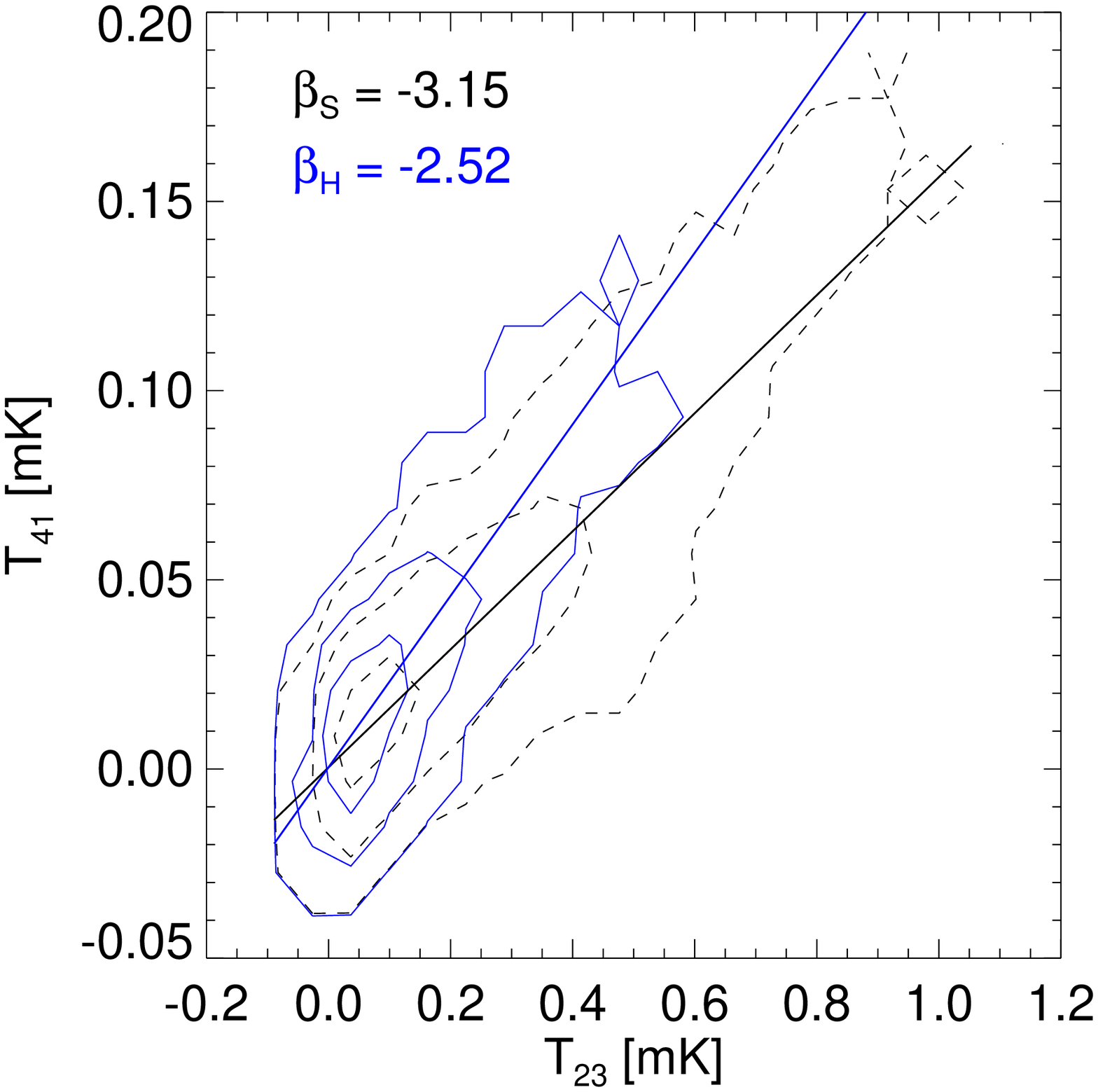}
}
\caption{
Number density contours of unmasked pixel (with $l=[-25:+25]$ and $b=[-45:0]$) 
temperatures in the $\Res_{H}$ (blue) and $\Res_{H+S}$ (dashed) maps for the RG8 
fits at 23, 33, and 41 GHz for CMB5.  Though the spectral slope is somewhat 
uncertain between each band and can vary significantly depending on which CMB 
estimator is used (see Table \ref{tbl:synchslope}), the best fit spectral 
slope for the haze emission is distinctly harder than for the total synchrotron.
}\label{fig:haze-scatter}
\epm

\bpm
\centerline{
  \includegraphics[width=0.9\textwidth]{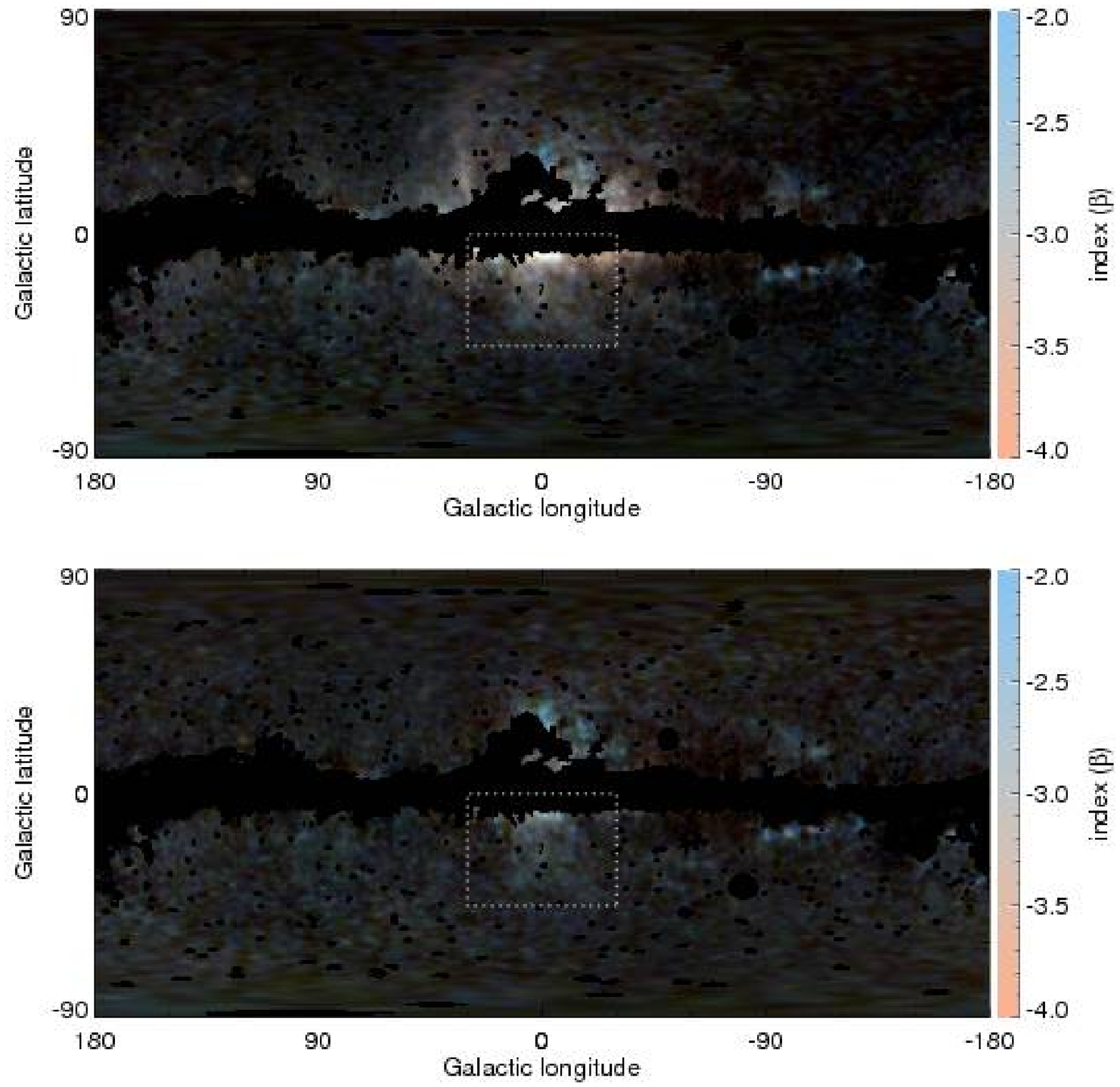}
}
\caption{
An RGB representation of $\Res_{H}$ and $\Res_{H+S}$ for RG8 with CMB5.  The color 
coding indicates the spectral index, in antenna temperature, 
of a given pixel.  In particular, the bluer 
haze region (\emph{box}) indicates a harder spectrum than the redder synchrotron emission.
}\label{fig:rgb-haze}
\epm

\reffig{haze-scatter} shows density contours for a scatter plot of the unmasked 
pixel values in $\Res_{H}$ and the total synchrotron residual map, $\Res_{H+S} = 
\Res_H + \rmbf{a}_{s} \rmbf{s}$, in antenna temperature for various frequency 
combinations.  The pixels shown have $l=[-25:+25]$ and $b=[-45:0]$.  Despite the 
large scatter, the $\Res_H$ emission appears to be a distinctly separate 
component of synchrotron emission with a spectral index that is significantly 
harder than the synchrotron (as shown in Table \ref{tbl:synchslope} this 
behavior persists for all CMB estimator types, though the precise spectral 
indices are very uncertain due to the CMB estimator bias describe in 
\refsec{ilc-bias}).  
This point is underscored in an RGB color coded 
map of the 23, 33, and 41 GHz $\Res_H$ and $\Res_S$ maps.  The haze emission is 
distinctly bluer (harder spectral index) than the total, redder (softer) 
synchrotron emission in the Galactic center.  

\begin{deluxetable}{|c|ccc|ccc|}
\tablehead{
CMB & & $\beta_{S}$ & & & $\beta_{H}$ & \\
estimator & 23/33 & 33/41 & 23/41 & 23/33 & 33/41 & 23/41
}
\startdata
 1 & -2.86 & -2.58 & -2.81 & -2.14 & -2.13 & -2.23 \\
 2 & -2.78 & -2.47 & -2.71 & -1.98 & -1.98 & -2.01 \\
 3 & -2.75 & -2.53 & -2.66 & -2.24 & -2.56 & -2.31 \\
 4 & -3.15 & -3.33 & -3.22 & -2.57 & -3.14 & -2.76 \\
 5 & -3.10 & -3.13 & -3.15 & -2.39 & -2.67 & -2.52 \\
 6 & -3.00 & -3.05 & -3.01 & -2.37 & -2.92 & -2.51
\enddata
\tablecomments{
The best fit spectral slopes for $\Res_{H+S}$ ($\beta_{S}$) and $\Res_{H}$ 
($\beta_{H}$) maps for the RG8 fits at 23, 33, and 41 GHz over unmasked pixels 
(with $l=[-25:+25]$ and $b=[-45:0]$, see \reffig{haze-scatter}).  Though there 
is significant scatter in the inferred spectral index, the haze residual is 
always \emph{harder} than the total synchrotron signal.
}\label{tbl:synchslope}
\end{deluxetable}

\reffig{inthaze} shows the total intensity of the haze as a function of distance 
south of the Galactic center.  The radial bins are 20 degrees wide and separated 
by 1 degree in latitude.  Our simple $1/r$ profile for the haze emission is not 
adequately describing the structure of the haze --- which is also clear from the 
residual maps in Figures \ref{fig:fullsky} and \ref{fig:gc}.  Furthermore, the 
systematic error bars due to chance correlation between the haze and the CMB are 
quite large, particularly at high frequencies.

\section{CMB estimator bias}
\label{sec:ilc-bias}

While the statistical uncertainty in our fits comes from measurement noise in the 
WMAP data, systematic uncertainties in our fits are dominated by contamination of 
the CMB estimator by the foreground components.  We have used six different CMB 
estimators, but they can be separated into two categories: ILC type estimators in 
which multiple WMAP bands are combined to approximately cancel the foregrounds 
(CMB1, CMB2, CMB3, and CMB5), and HF estimators in which a model for the thermal 
dust and free--free emission is removed from the highest WMAP band to 
approximately leave the CMB (CMB4).

\subsection{ILC estimators}

\citet{hinshaw07} point out that the WMAP data consists of the ``true'' CMB, 
$T_c$ plus some additional contamination by foreground components $T_j$ which have 
some spectral dependence $N_{i,j}$ where $i$ and $j$ represent the observing band 
and foreground component respectively.  That is,
\be
  W_i = T_c + \sum_j N_{i,j} T_j
\ee
so that an internal linear combination of the WMAP bands is,
\bel{ilcdef}
  \begin{array}{ccl}
    L & = & \sum_i \zeta_i W_i \\
&& \\
      & = & \sum_i \zeta_i T_c + \sum_{j,i} \zeta_i N_{i,j} T_j \\
&& \\
      & = & T_c + \sum_j \Gamma_j T_j,
  \end{array}
\ee
where $\zeta_i$ are constrained to sum to one in order to preserve
unity response to the CMB, and $\Gamma_j$ parameterizes the
contamination of $L$ by foreground $j$. 

\bpm
\centerline{
  \includegraphics[width=0.35\textwidth]{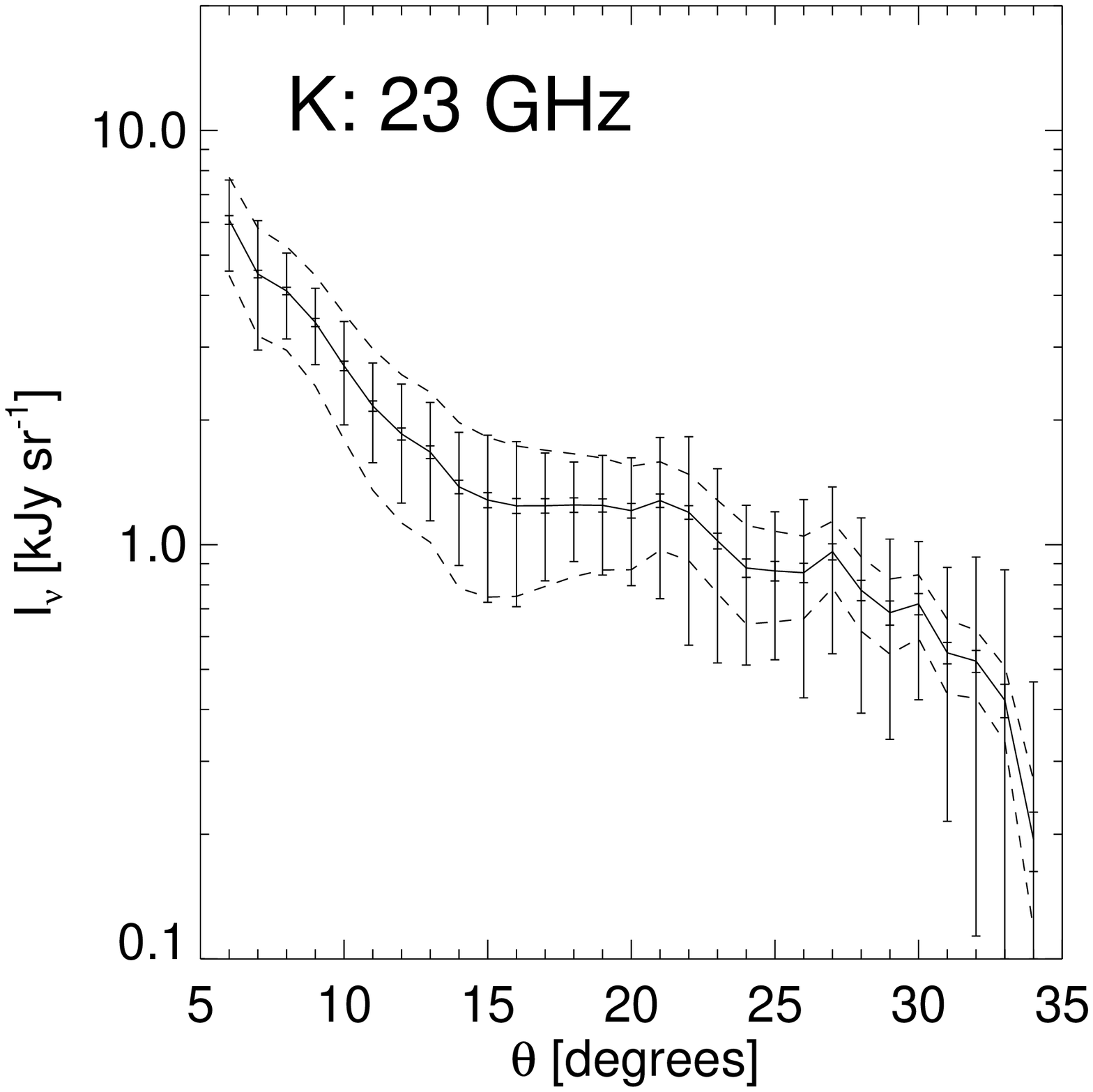}
  \includegraphics[width=0.35\textwidth]{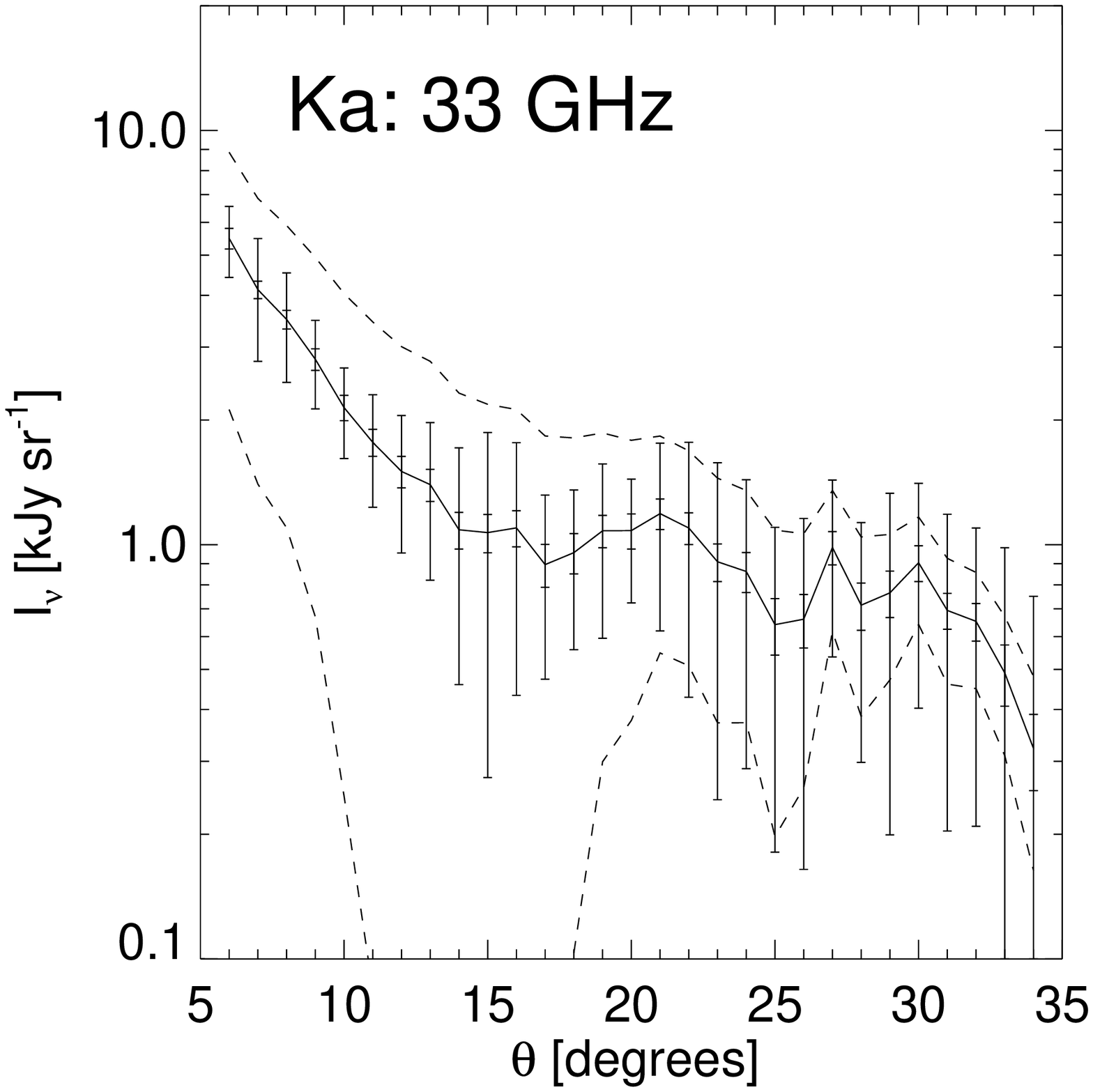}
}
\centerline{
  \includegraphics[width=0.35\textwidth]{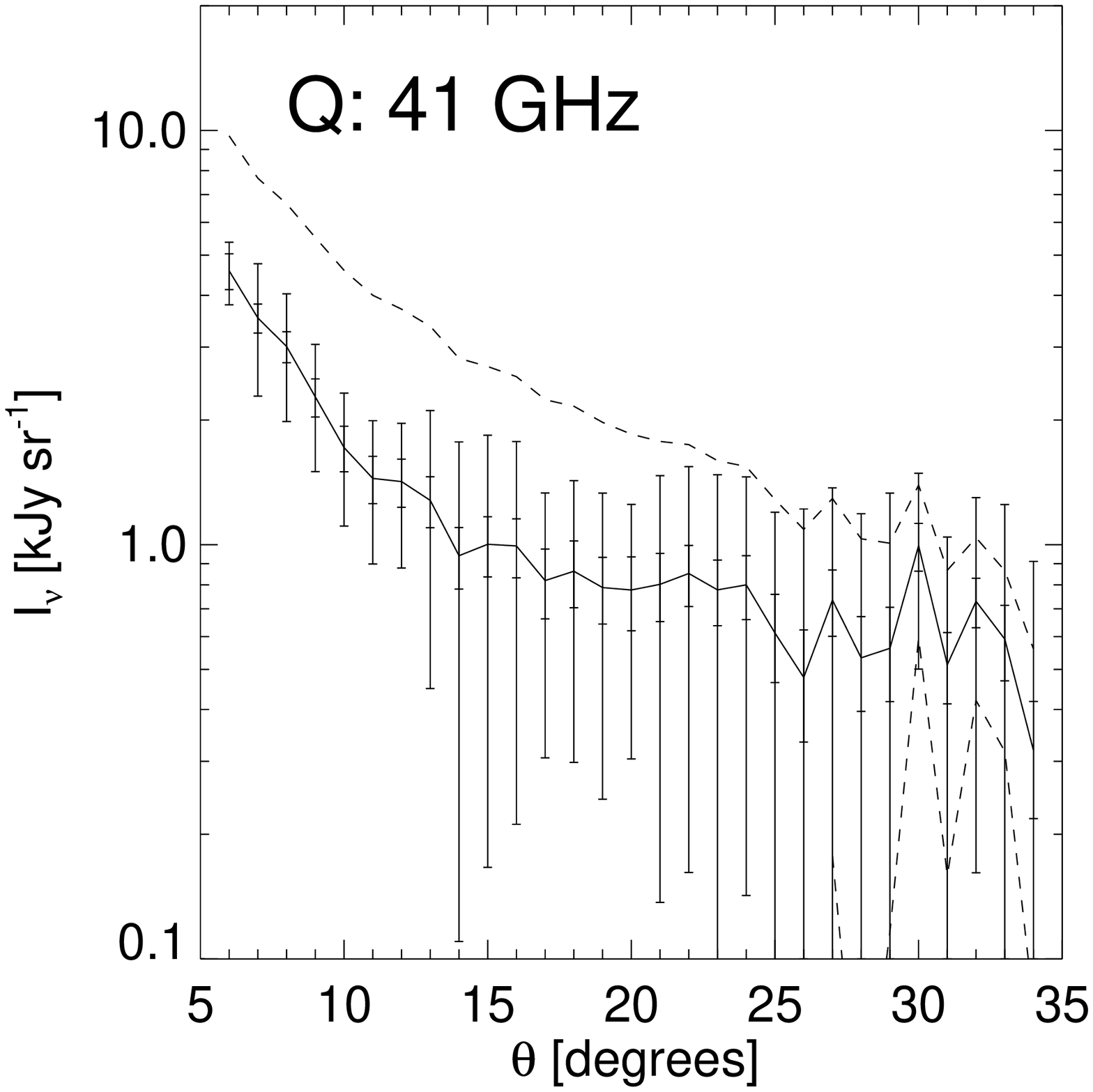}
  \includegraphics[width=0.35\textwidth]{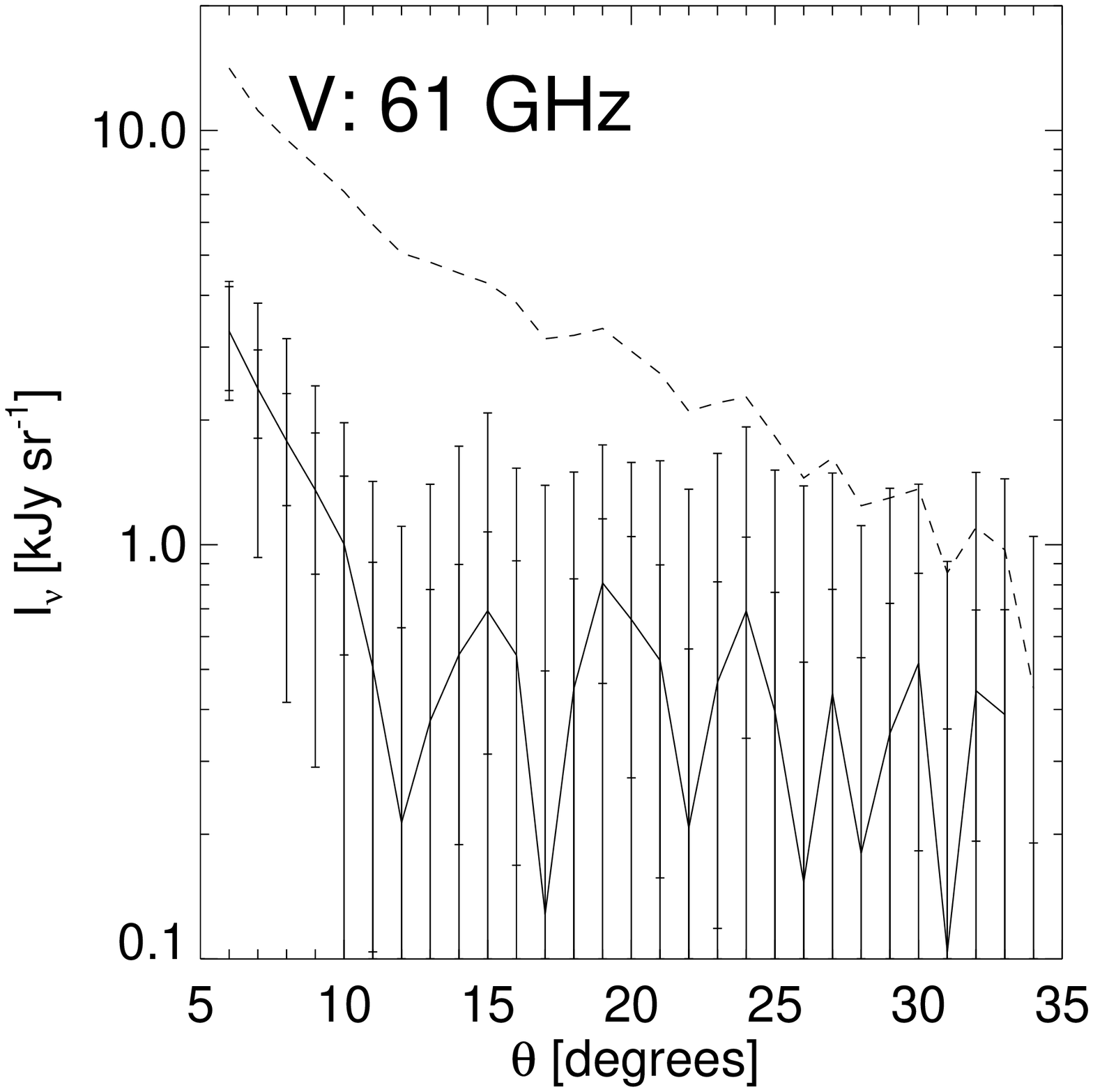}
}
\caption{
Integrated haze (from the bottom panel of \reffig{fullsky-regions}) in kJy/sr as a 
function of radial distance south of the Galactic center.  The radial bins are 20 
degrees wide and separated by 1 degree in longitude.  The inner error bars are due 
to the formal error on the fit coefficients, the outer error bars are the 
1-$\sigma$ standard deviation of the temperature fluctuations in a given radial 
bin, and the dotted lines represent the bias due to chance correlation between the 
CMB and the haze.
}\label{fig:inthaze}
\epm

Let us consider that there are four types of foreground emission: free--free 
($F$), dust ($D$), soft synchrotron ($S$), and the haze ($H$).  Each of these has 
an associated $\Gamma$, and so the ILC is given by
\be
  L = T_{c} + \Gamma_{F} T_{F} + \Gamma_{D} T_{D} 
                      + \Gamma_{S} T_{S} + \Gamma_{H} T_{H}.
\ee
Since $\zeta_i$ are chosen to minimize the variance in $L$, 
\bel{diffeq}
  \frac{\partial \langle L^2 \rangle}{\partial \Gamma_j} = 0
\ee
for all $j$, and we have explicitly assumed mean subtracted maps so that $\langle 
L \rangle = \langle T_{c} \rangle = \langle T_{j} \rangle = 0$.  Now,
\bel{ilcsq}
  \begin{array}{ccl}
  L^2 & = & T_{c}^2 + 2\sum_{j} \Gamma_j T_{c} T_{j} \\
 & & \\

                &   & + 2 \sum_{j} \sum_{k \neq j} \Gamma_j \Gamma_k T_{k} T_{j} 
\\
 & & \\
                &   & + \sum_{j} \Gamma_j^2 T_{j}^2
  \end{array}
\ee
so that \refeq{diffeq} reads
\bel{long}
  0 = 2 \langle T_{c} T_{j} \rangle + 
      2 \sum_{k \neq j} \Gamma_{k} \langle T_{j} T_{k} \rangle +
      2 \Gamma_j \langle T_{j}^2 \rangle.
\ee
(note: the notation in \refeq{long} can be compactified into $0=\sum_{\psi} 
\Gamma_{\psi} \langle T_{j} T_{\psi} \rangle$ where $\psi=[c,F,D,S,H]$ and 
$\Gamma_c \equiv 1$)  Thus, we can solve explicitly for
\bel{gammaj}
  \Gamma_{j} = - \frac{\langle T_{c} T_{j} \rangle 
                     + \sum_{k \neq j} \Gamma_k \langle T_{j}T_{k} \rangle}
                    { \langle T_{j}^2 \rangle}
\ee
In the limit of only one foreground $T_{f}$ the second term (which represents 
cross correlation between the different foreground morphologies) disappears and 
this reduces to $\Gamma = \Gamma(T_{c},T_{f}) = -\sigma_{cf}/\sigma_{f}^2$ as 
derived with a similar method by Hinshaw et al (2006).

According to \refeq{gammaj}, the ILC map is biased towards anti-correlation 
between different ``true'' foreground emission morphologies and the ``true'' 
CMB.  However, since we do not know the true $T_{c}$ or $T_{j}$ \emph{a priori}, 
we have no information about how much of each foreground to add back into $L$ in 
order to correct for this factor.

We can use Monte Carlo techniques to estimate the amplitude (but \emph{not} the 
sign) of the error in $L$ due to foreground contamination as follows.  We 
construct 100 realizations of the CMB by generating random phases for the 
measured power, $C_{\ell}$, in each Fourier mode of the binned three year 
WMAP power spectrum ($\ell_{\rm max} = 986$) from \citet{spergel07}.  For each 
realization $T'_{c}$ we can estimate $\Gamma$ with a reasonable foreground 
template $T'_{j}$ which we suppose traces the morphology of the emission 
(see \refsec{foremaps}).

\bpm
\centerline{
  \includegraphics[width=0.35\textwidth]{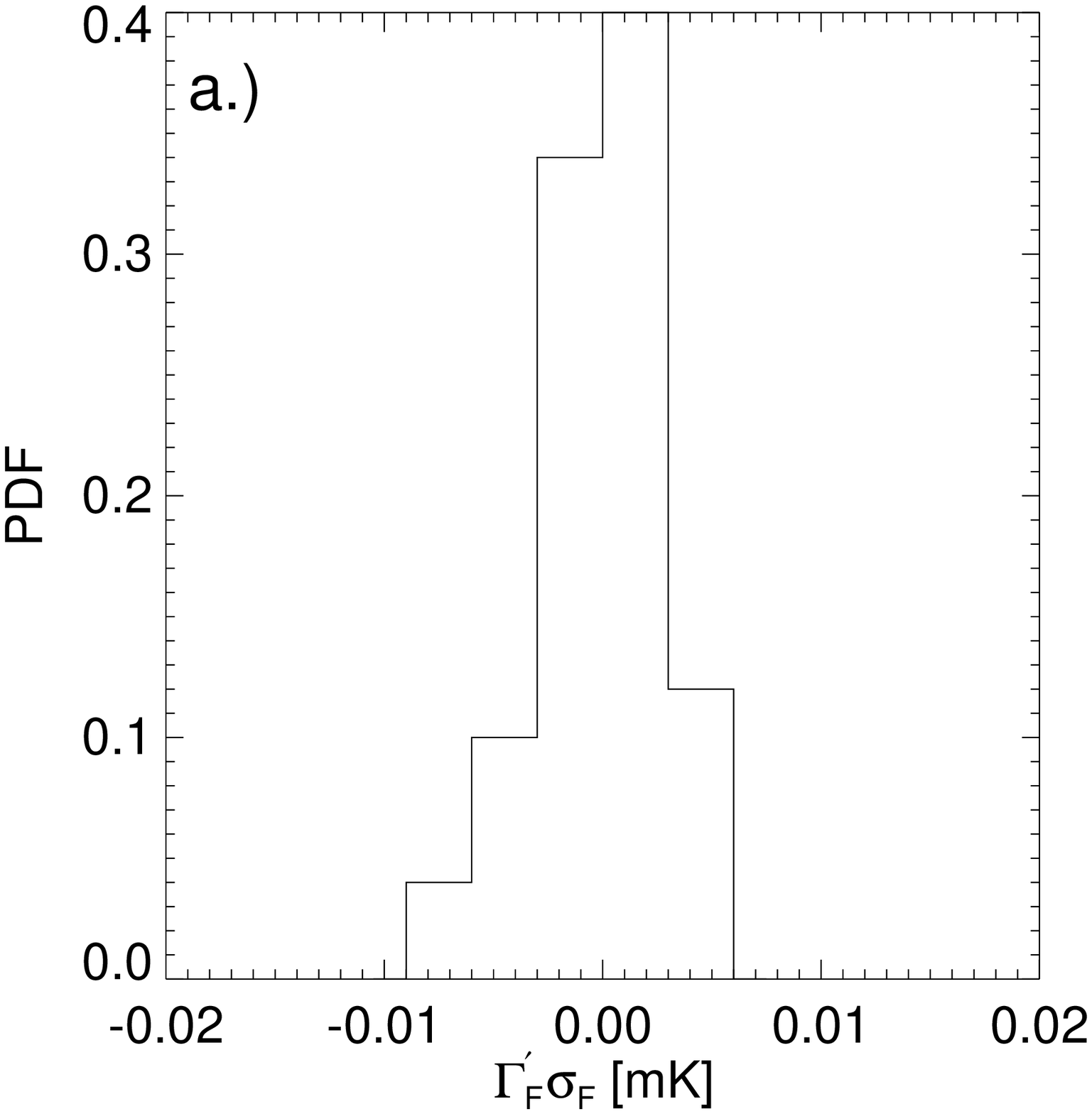}
  \includegraphics[width=0.35\textwidth]{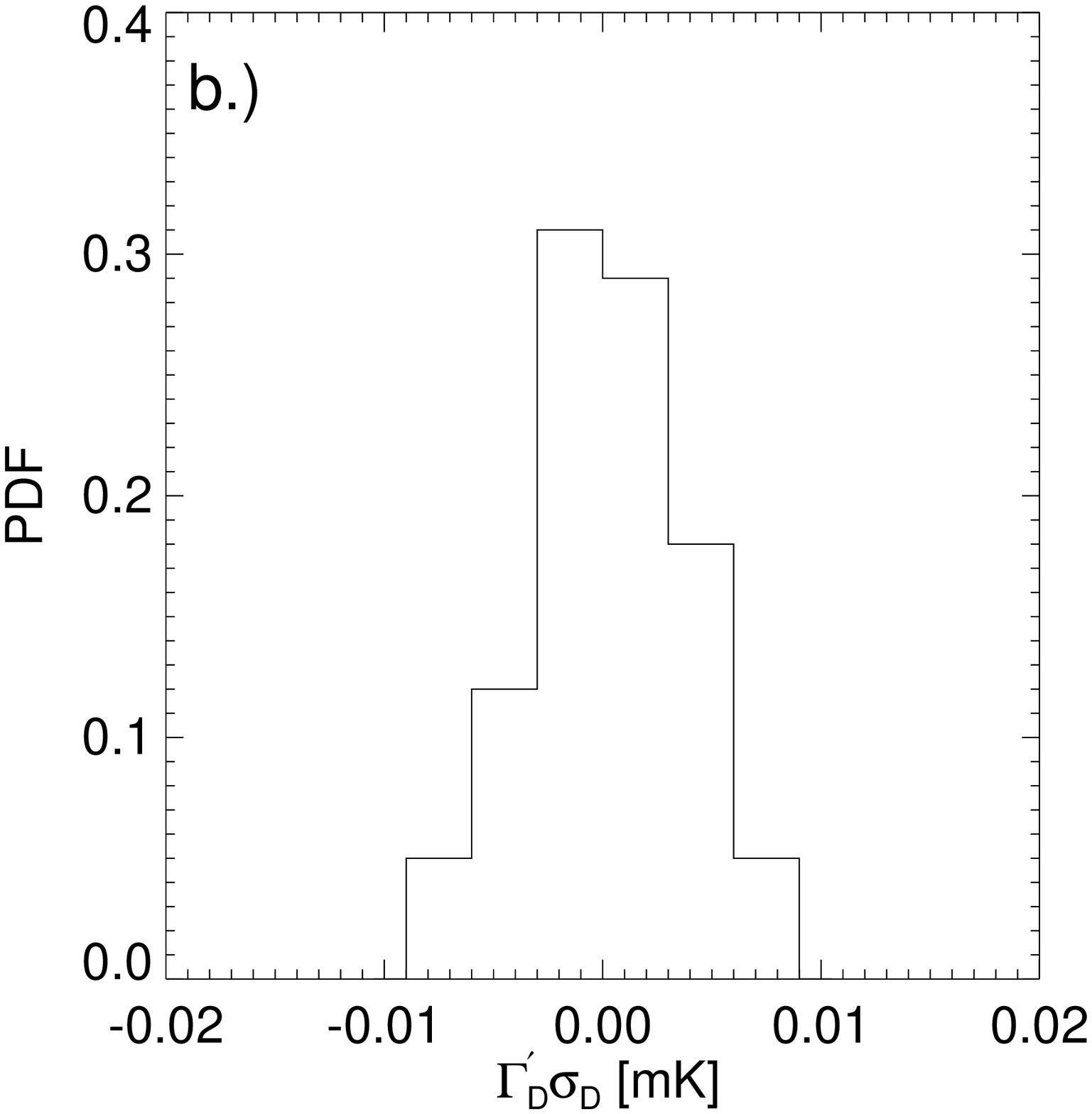}
}
\centerline{
  \includegraphics[width=0.35\textwidth]{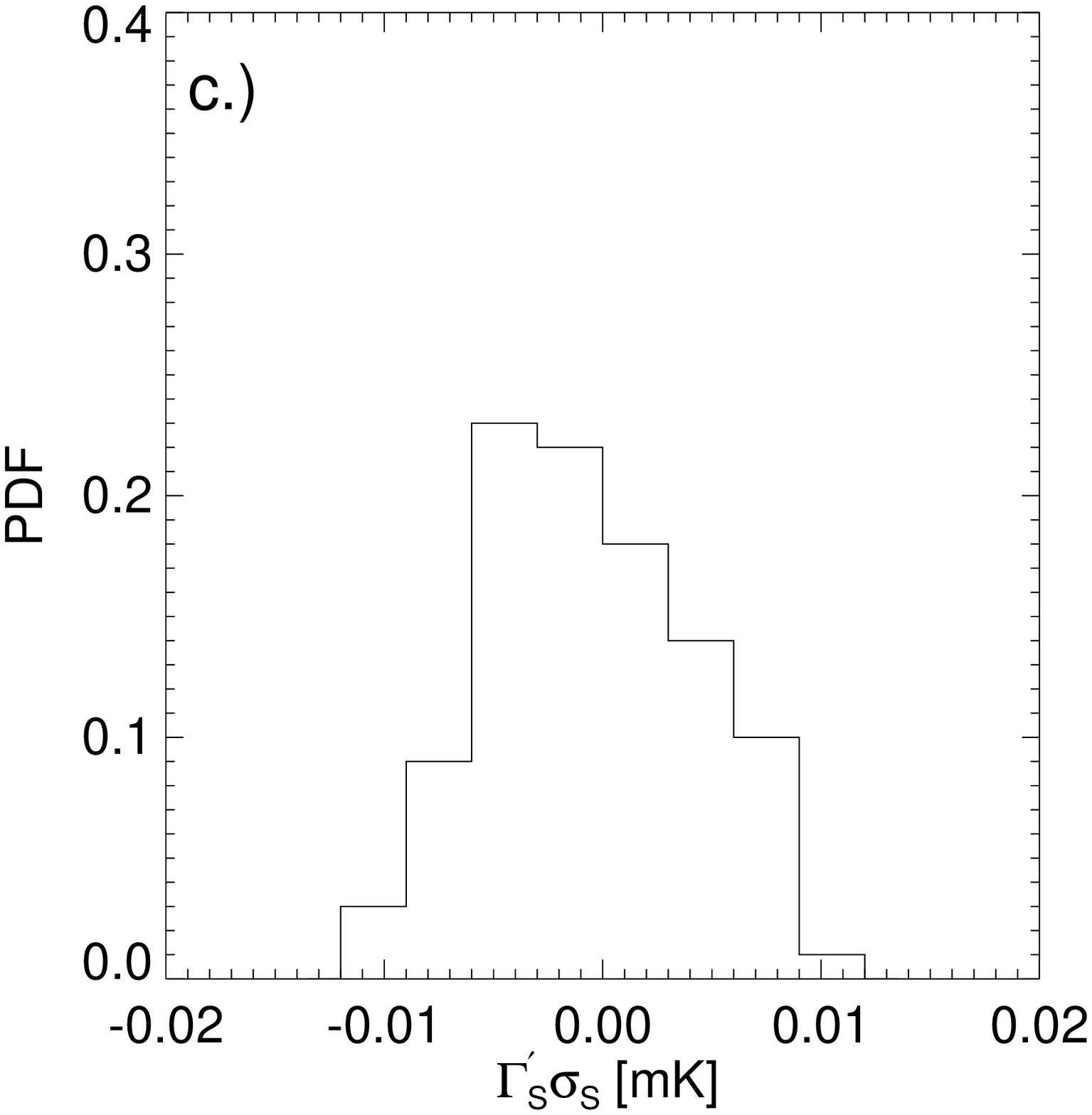}
  \includegraphics[width=0.35\textwidth]{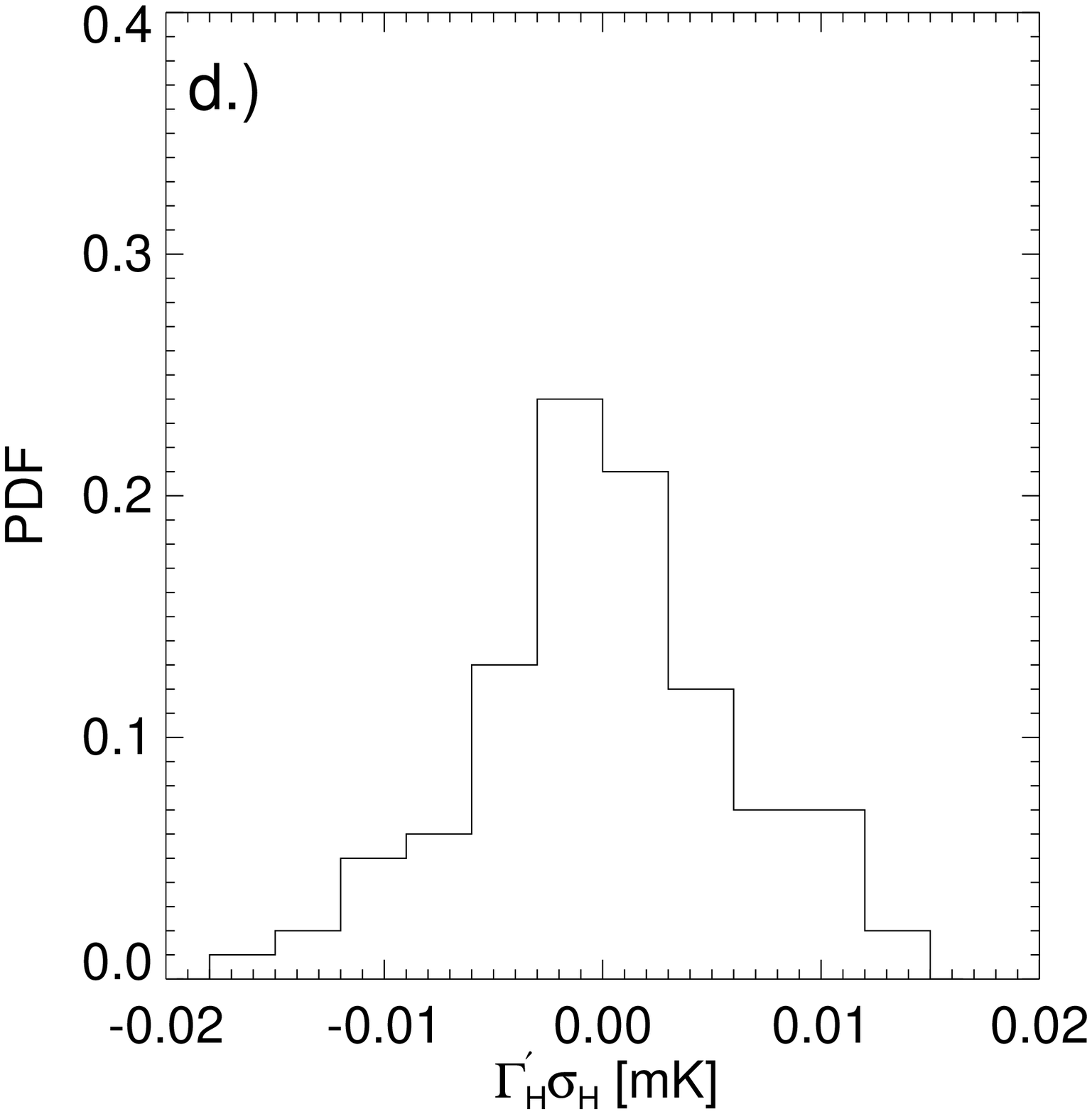}
}
\caption{
Distribution of ILC chance cross correlation biases for 100 realizations of the 
CMB sky.  a.) H$\alpha$-correlated emission, b.) dust-correlated emission, c.) 
Haslam-correlated emission, and d.) haze emission.
}\label{fig:ilc-bias}
\epm

Figure \ref{fig:ilc-bias} shows histograms of 
\be
  \Gamma' \sigma_j = \Gamma(T'_{c},T'_{j}) \sigma_j,
\ee
where $\sigma_j$ is the standard deviation over unmasked pixels of the foreground 
template $T'_{j}$ ($T'_{j} = $ H$\alpha$, FDS, Haslam, and 
haze), for all 100 realizations of the CMB sky.  We multiply by $\sigma_j$ to 
obtain the contamination in a ``typical'' pixel in the ILC map.  Each histogram 
has a mean consistent with zero,
\be
  \langle \Gamma' \rangle \approx 0 \mbox{ for each } T'_{j}
\ee
as expected since the CMB realization $T'_{c}$ is just as likely as $-T'_{c}$.  
The implication is that we can only estimate the \emph{uncertainty} in $T_{\rm 
ILC}$, we cannot explicitly correct for the bias.  

It is important to note that, although $L$ cannot be corrected for the
cross-correlation bias, the net effect is to \emph{decrease} the
variance relative to the true CMB, $T_c$.  Since the bias of the
variance is not mean zero, the effect on the power spectrum of the CMB
can be estimated \citep[cf.][]{hinshaw07}.

\subsection{HF type estimators}
The bias in a HF estimator is easy to understand and due entirely to the 
residual foreground emission after subtraction of thermal dust and free--free.  
Assuming perfect subtraction of these two foregrounds, whatever other 
foregrounds are left in the HF estimator are subtracted off all lower frequency 
bands leading to a systematic bias towards softer spectra for the soft 
synchrotron and haze components.

\subsection{Forecasts for \emph{Planck}}
\label{sec:planck}

The principle advantage of the \emph{Planck} mission over WMAP is the large range of 
frequency coverage.  In particular, the multiple channels at very high frequency 
reduce the foreground problem to (mostly) a single emission mechanism, thermal 
dust.  In order to exploit this feature, we suggest a new approach based on an 
ILC type CMB estimator.  However, instead of minimizing the variance (which we 
have seen leads to large cross-correlation uncertainties), we choose the ILC 
coefficients $\zeta_i$ to cancel as many \emph{power law} foreground components 
as possible.  This estimator can then be subtracted off of lower frequency bands 
with very minimal bias contamination of the inferred foreground spectra

\emph{Planck}'s frequency coverage is 30, 44, 70, 100, 143, 217, 353, 545, and 857 GHz.  
If we estimate that spinning dust emission is negligible above 100 GHz, then 
these bands contain mostly thermal dust ($T \propto \nu^{\beta_+}$ with $1.6 
\leq \beta_+ \leq 2.3$) and also small amounts of free--free, synchrotron, and 
haze ($T \propto \nu^{\beta_-}$ with $-3.1 \leq \beta_- \leq -2.1$).

In order to find an ILC that optimally cancels out this range of power 
law indices, we formulate the error function,
\bel{delta}
  \delta_{94}(\beta) = \frac{T_{94}}{\sum_i \zeta_i T_i},
\ee
for a foreground spectrum, normalized to 94 GHz, $T_i = T_{94} (\nu_i/94 {\rm  \ 
GHz})^{\beta}$, where $\nu_i$ are the \emph{Planck} bands above 100 GHz and the sum is over 
bands.  The physical interpretation of $\delta_{94}(\beta)$ is that it 
represents the fractional bias in the inferred 94 GHz amplitude of a foreground 
with power law index $\beta$.  We minimize $\int \delta_{94}^2 d\beta$ with 
respect to $\zeta_i$ over the range $\beta_+$ and $\beta_-$ defined above.  
\reffig{pl-spec} shows the resultant $\delta_{94}$.  With these $\zeta_i$ we can 
form the \emph{Planck} ILC $L_{\rm P} = \sum_i \zeta_i P_i$, where $P_i$ are the \emph{Planck} 
data in thermodynamic mK at band $i$.  We find that 
\bel{plc-ilc}
\begin{array}{ccl}
  L_{\rm P} $ = $ -1.49 P_{143} + 3.21 P_{217}  - 0.74 P_{353} \\
				$ $ + 0.02 P_{545} - 3.21 \times 10^{-5} P_{857}
\end{array}
\ee
minimizes the area under the $\delta_{94}^2$ curve over the $\beta$
range of interest.  Using only the high frequency channels in this way
avoids ambiguities due to an uncertain spinning dust spectrum.

To estimate the contamination of $L_{\rm P}$ by synchrotron and haze emission 
we must make a couple of assumptions.  First, let us assume that these 
foregrounds follow a power law with with amplitudes of roughly our fit results 
at 23 GHz from \reffig{spectra-fs} and indices $-3.1 \leq \beta_S \leq -2.7$ and 
$-2.7 \leq \beta_H \leq -2.4$ respectively.  Second, we make the implicit 
approximation that the morphologies of the foreground emission mechanisms do not 
change with frequency and that they are still well represented by the templates 
in \refsec{foremaps}.

\reffig{pl-spec} shows our estimates for the error bars on the ``mock true'' 
soft synchrotron and haze spectra at WMAP frequencies given three CMB 
estimators: a WMAP only ILC (leading to cross correlation bias errors), our WMAP 
HF estimator CMB4, and the $L_{\rm P}$ estimator.  As we have already seen, the 
WMAP only ILC clearly yields very large uncertainties, though they have the 
advantage of being mean zero (averaged over an ensemble of CMB realizations).  
CMB4 does significantly better, though now the errors are systematically biased 
towards softer spectra.  With $L_{\rm P}$, the bias errors are almost completely 
eliminated.  For a foreground with indices $\beta =$ -3.1, -2.7, and -2.4, the 
bias errors at 23 (94) GHz are $-$0.11\%, $-$0.06\%, and $+$0.13\% ($-$7.6\%, 
$-$2.2\%, and $+$3.3\%) respectively.

We emphasize that we have by no means attempted to formulate the ``optimal'' 
foreground removal algorithm for \emph{Planck}.  There have been numerous 
discussions on the topic describing and comparing different cleaning 
methods.\footnote{E.g., 
\citet{brandt94} undertook an early study of foreground removal with multiple 
frequency bands, \citet{tegmark03} and \citet{deO06} describe an ILC type 
method in which the weighting depends on the multipole $\ell$ of the spherical 
harmonic decompostion, \citet{barreiro04} expands on the MEM analysis,
\citet{eriksen06} describe a component separation method, \citet{hansen06} 
tackle the problem with a wavelet analysis, and \citet{dodelson97} attempts to 
``design'' an experiment with optimal frequency coverage for removing 
foregrounds.}
Rather, we have demonstrated that, by exploiting the many bands and high 
frequency coverage, this simple \emph{Planck} ILC will substantially reduce the 
systematic biases in the inferred foreground spectra compared to WMAP.

\bpm
\centerline{
  \includegraphics[width=0.3\textwidth]{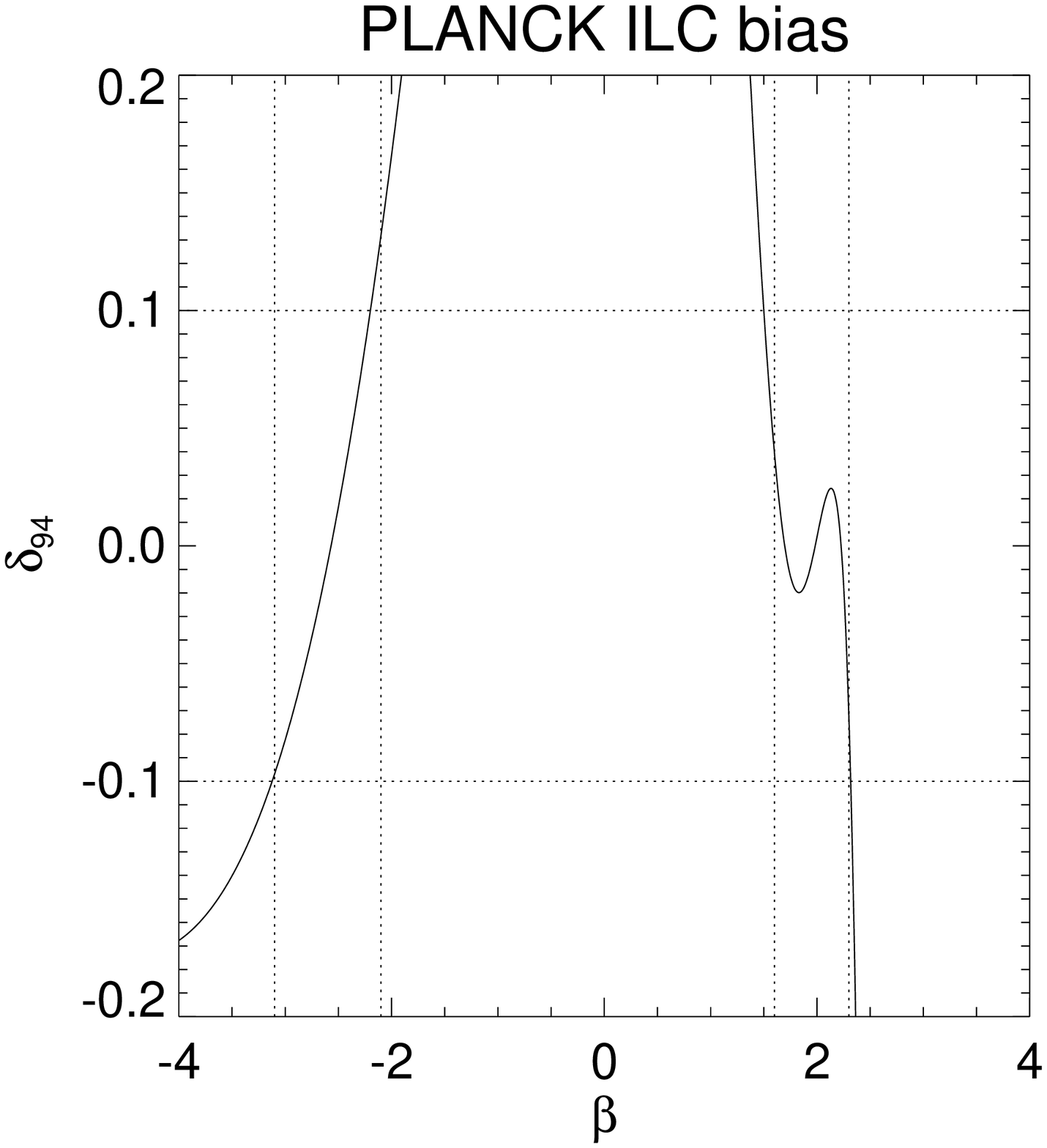}
  \includegraphics[width=0.3\textwidth]{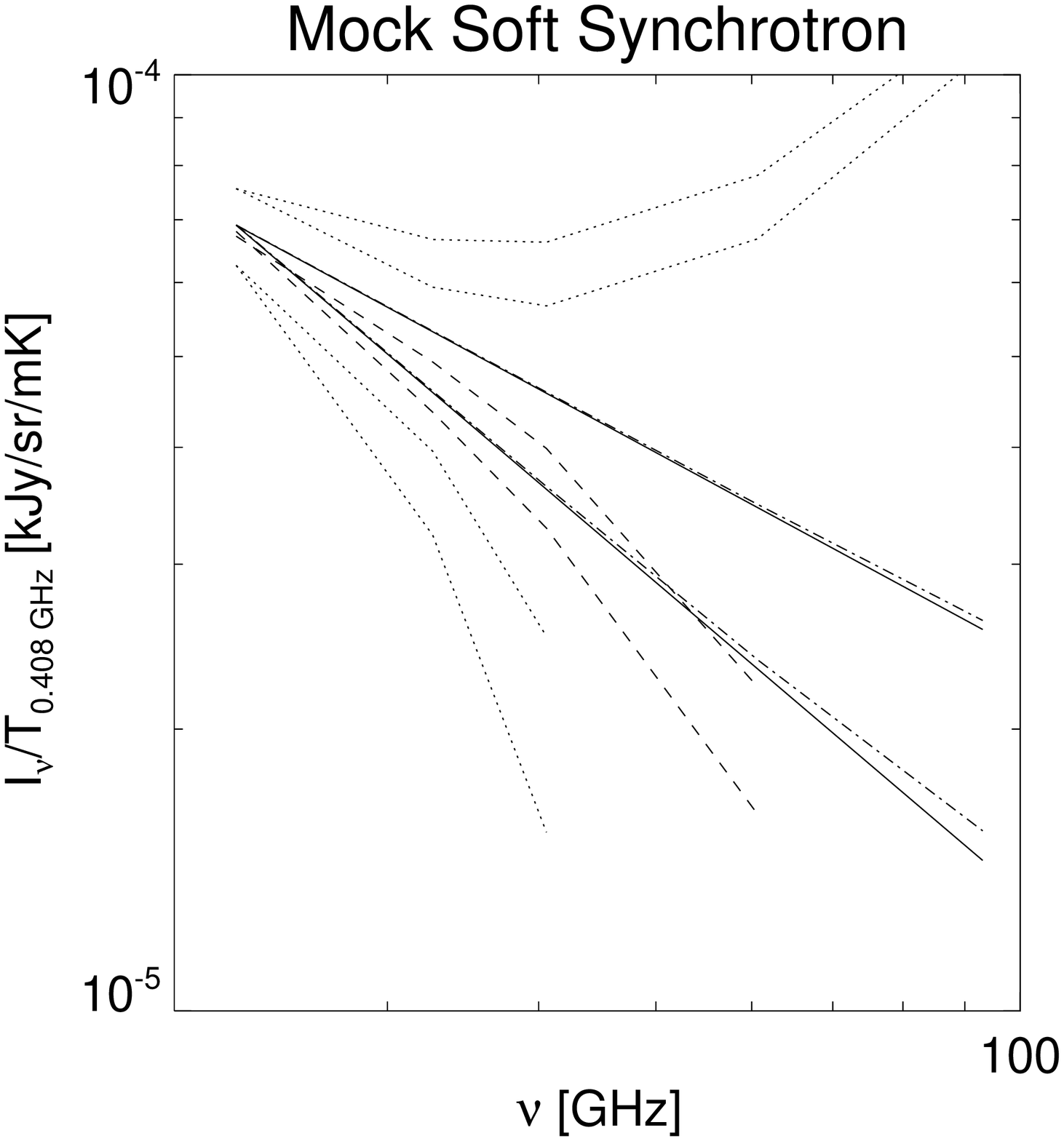}
  \includegraphics[width=0.3\textwidth]{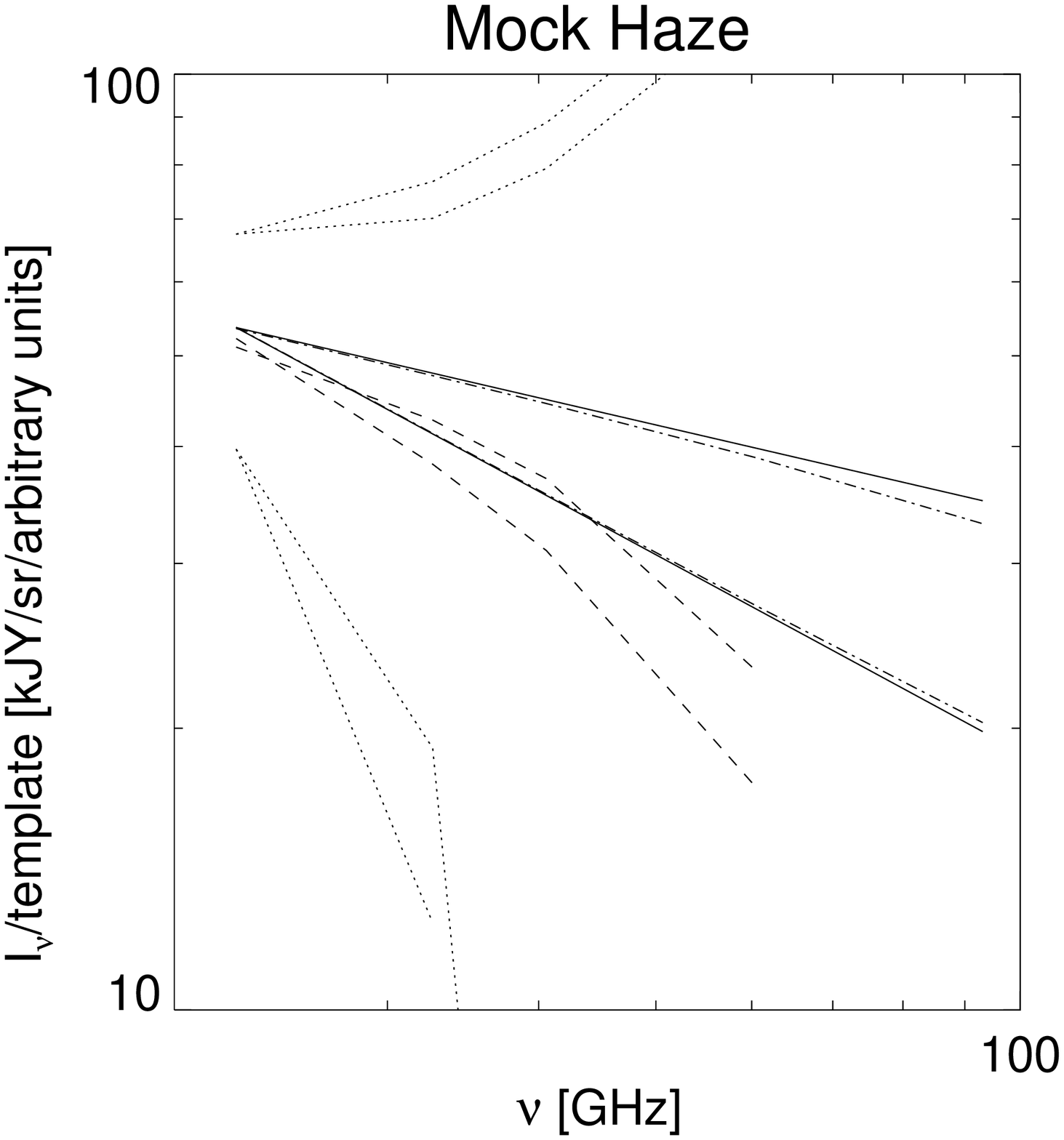}
}
\caption{
Left panel: fractional error in the inferred 94 GHz amplitude of a foreground 
with $T \propto \nu^{\beta}$ due to contamination of the \emph{Planck} ILC $L_{\rm P}$ 
(see \refeq{plc-ilc}).  Center and right panels: biases due to foreground 
contamination of different CMB estimators for mock soft synchrotron ($T_S 
\propto \nu^{\beta_S}$ with $-3.1 \leq \beta_S \leq -2.7$) and the haze ($T_H 
\propto \nu^{\beta_H}$ with $-2.7 \leq \beta_H \leq -2.4$).  The mock spectrum is 
shown with a solid line.  The dotted lines are the mean zero ILC bias errors 
using only WMAP data alone, the dashed lines are the systematic biases due to 
CMB4, and the dot-dashed lines are the systematic bias using $L_{\rm P}$.  The 
systematic biases with \emph{Planck} are more than an order of magnitude smaller than 
for a WMAP HF estimator. 
}\label{fig:pl-spec}
\epm

\section{Discussion}
We have carried out a multi-linear regression fit to the WMAP data using 
foreground templates for free--free, soft synchrotron, and dust emission.  We 
perform fits over both the (nearly) full-sky as well as smaller regions of 
interest.  Our method simultaneously fits the amplitude and spectrum of each 
emission component and is immune to priors or initial ``guesses'' for the 
template amplitudes.

Importantly, we find that the spectra of the
foreground emission mechanisms cannot be determined to high accuracy
with the WMAP data.  The root cause is that any estimator of the CMB
will necessarily be contaminated by the foregrounds to some degree, so
that subtracting the CMB estimator from the WMAP data systematically
biases the inferred foreground spectra.  Thus, the fit spectrum for
each foreground is contaminated by some amount of CMB spectrum.  The
degree of systematic bias in the foreground spectra can be quite large
and varies dramatically among our different CMB estimators.  It is
worth noting that the degree of contamination of a given CMB estimator
does not significantly affect the variance of the estimator.

We find that, upon removing the free--free, dust, and soft synchrotron emission, 
the 3--year WMAP data still contains the ``haze'' seen by Finkbeiner (2004).  
Since this haze emission is present in all residual maps regardless of which CMB 
estimator is used, we conclude that it is not an artifact of imperfect CMB 
subtraction.  We have included a simple $1/r$ template for the haze to relax the 
stress on the other foreground components in our fit.  This template imperfectly 
matches the morphology of the haze, but allows us to simultaneously fit an 
approximate amplitude and spectra of the haze emission as well.

Despite the above issue of bias errors, we can make the following concrete 
statements about our fit foreground spectra.

\begin{enumerate}
\item

We find that the H$\alpha$-correlated emission does not follow a simple 
$I_{\nu} \propto \nu^{-0.15}$ free--free powerlaw as expected from theory.  Rather 
we find a bump in the H$\alpha$-correlated emission which we argue in our 
companion paper D+F (2007) can be explained by a mixture of free--free gas and 
spinning dust.  We emphasize that this bump is \emph{not} due to the contamination 
of the CMB estimator by the foregrounds.

\item
The dust-correlated emission has the now familiar fall off from 23 to 41 GHz 
consistent with a superposition of D+L spinning dust spectra and then a rise at 94 
GHz from thermal dust emission.  We cannot use our multi-region fits to map out 
variations in the spinning dust spectrum from place to place across the sky since 
the uncertainty from the systematic bias dominates.

\item
Our soft synchrotron spectra are very sensitively dependent on which CMB estimator 
is used.  For example, with the WMAP team's published ILC, the soft synchrotron 
spectrum actually turns up at high frequencies, while other estimators show a 
steadily falling spectrum (with $T_{\nu} \propto \nu^{\beta}$ where $\beta \approx 
-3.0$.)  This ambiguity is entirely due to the contamination of the CMB 
estimator by soft synchrotron type emission.

\item The haze spectrum is similarly very uncertain due to CMB
  estimator bias, however, for a given estimator, the haze spectrum is
  always harder than the soft synchrotron.  For reasons outlined in
  \refsec{haze-vs-sync}, we suspect that the haze emission is due to
  hard synchrotron from a separate component of very energetic
  electrons near the Galactic center.  Possible sources include
  products of dark matter annihilation \citep{hooper07} or a single
  energetic event in the last million years, e.g. a gamma-ray burst
  \citep{broderick07}.
\end{enumerate}

Upcoming experiments like \emph{Planck} will significantly reduce the
problem of CMB estimator bias in the inferred foreground spectra due
to the improved high frequency coverage.  In particular, for
frequencies $\nu > 100$ GHz the \emph{Planck} data will be virtually free of
all foregrounds except for thermal dust.  Removing this one bright
component with five bands is significantly less ambiguous than
removing three or four bright components with five bands.
Furthermore, what little free--free, soft synchrotron, and haze emission there 
is at these high frequencies can be roughly eliminated in a \emph{Planck} CMB estimator 
through the appropriate choice of ILC coefficients.  We estimate that the 
systematic uncertainties in soft synchrotron and haze spectra will be reduced by 
more than an order of magnitude compared to WMAP.

Galactic foreground emission represents the most significant contaminant in 
determining cosmological parameters.  Ironically, the reverse statement is also 
true and in order to make progress towards determining the ``true'' spectra of 
the foreground emission, a more reliable estimate of the CMB sky is required.

\acknowledgements
We thank Gary Hinshaw, Angelica de Oliveira-Costa, Max Tegmark, and the 
anonymous referee for helpful comments and suggestions.  Some of the results in 
this paper were derived using HEALPix \citep{gorski99,calabretta07}.  This 
research made use of the NASA Astrophysics Data System (ADS) and the IDL 
Astronomy User's Library at Goddard\footnote{Available at 
\texttt{http://idlastro.gsfc.nasa.gov}}.  DPF and GD are supported in part by 
NASA LTSA grant NAG5-12972.

\end{document}